# Are the Moon's nearside-farside asymmetries the result of a giant impact?

Meng-Hua Zhu[1,2,6*], Kai Wünnemann[2], Ross W. K. Potter[3], Thorsten Kleine[4], Alessandro Morbidelli[5]

[1] Space Science Institute, Macau University of Science and Technology, Avenue Wailong, Taipa, Macau

[2] Museum für Naturkunde, Leibniz Institute for Evolution and Biodiversity Science, Berlin, Germany

[3] Department of Earth, Environmental and Planetary Sciences, Brown University, Providence, RI, USA

[4] Institut für Planetologie, University of Münster, Münster, Germany

[5] Département Lagrange, University of Nice–Sophia Antipolis, CNRS, Observatoire de la Côte d'Azur, Nice, France

[6] CAS Center for Excellence in Comparative Planetology, China

*Correspondence to: mhzhu@must.edu.mo;   853-8897-2024

**Abstract**

The Moon exhibits striking geological asymmetries in elevation, crustal thickness, and composition between its nearside and farside. Although several scenarios have been proposed to explain these asymmetries, their origin remains debated. Recent remote sensing observations suggest that (1) the crust on the farside highlands consists of two layers: a primary anorthositic layer with thickness of ∼30-50 km and on top a more mafic-rich layer ∼ 10 km thick; and (2) the nearside exhibits a large area of low-Ca pyroxene that has been interpreted to have an impact origin. These observations support the idea that the lunar nearside-farside asymmetries may be the result of a giant impact. Here, using quantitative numerical modeling, we test the hypothesis that a giant impact on the early Moon can explain the striking differences in elevation, crustal thickness, and composition between the nearside and farside of the Moon. We find that a large impactor, impacting the current nearside with a low velocity, can form a mega-basin and reproduce the characteristics of the crustal asymmetry and structures comparable to those observed on the current Moon, including the nearside lowlands and the farside's mafic-rich layer on top of a primordial anorthositic crust. Our model shows that the excavated deep-seated KREEP (potassium, rare-earth elements, and phosphorus) material, deposited close to the basin rim, slumps back into the basin and covers the entire basin floor; subsequent large impacts can transport the shallow KREEP material to the surface, resulting in its observed distribution. In addition, our model suggests that prior to the asymmetry-forming impact, the Moon may have had an $^{182}$W anomaly compared to the immediate post-giant impact Earth's mantle, as predicted if the Moon was created through a giant collision with the proto-Earth.

**Plain Language Summary**

Beginning with the Apollo era, spacecraft observations have shown that the Moon has striking asymmetries between its nearside and farside in topography, crustal thickness, and composition. These asymmetries are likely a product of very early geological processes on the Moon. Understanding their formation mechanism may help to constrain models of global lunar evolution and magma-ocean crystallization. Several hypotheses have been suggested, though none can explain the observations satisfactorily.

Recent spacecraft observations from the Gravity Recovery and Interior Laboratory (GRAIL) mission indicate that the farside crust is ~ 20-km thicker than the nearside crust, and remote sensing data shows that this extra crust is composed of a mafic-rich layer covering the primary crust. This layered crustal structure on the farside, together with a large area of low-Ca pyroxene on the nearside observed by Kaguya mission that was explained to be formed via impact through melting a mixture of crust and mantle materials, means that a giant impact on the nearside may explain the nearside-farside asymmetries. To investigate this possibility, we quantitatively studied the giant impact theory using numerical modeling. Our models confirm that a giant impact on the nearside can explain nearside-farside asymmetries.

We demonstrate that a large body slowly impacting the nearside of the Moon can reproduce the observed crustal thickness asymmetry and form both the farside highlands and the nearside lowlands. Additionally, the model shows that the resulting impact ejecta would cover the primordial anorthositic crust to form a two-layer crust on the farside, as observed. Overall, the modeling results are generally in agreement with assumptions that are based observations and provide credible explanations for the observed asymmetries in crustal thickness and elevation. This work also provides a plausible explanation for the existence of KREEP (potassium, rare-earth element, and phosphorus) on the lunar surface. A very important implication of this work is that it can explain the conundrum about isotopic differences between the Earth and Moon, particularly the significant anomaly of $^{182}$W in the Moon, as this anomaly would occur if this giant impact added material to the Moon after the initial Moon-forming. Our model can thus explain this isotope anomaly in the context of the giant impact scenario of the Moon's origin.

In summary, this work quantitatively supports the long-standing hypothesis that a giant impact resulted in the Moon's nearside-farside asymmetries and the Procellarum KREEP terrain was formed as a consequence of such an impact event. In addition, this work also provides a reference for reconstructing the early history of planetary bodies with similar asymmetries, such as Mars.

# 1. Introduction

Since the Apollo era, it has been known that the Moon has striking nearside-farside asymmetries in topography, crustal thickness, and chemical composition (*Toksoz et al, 1974; Zuber et al., 1994; Lawrence et al., 1998*). On the nearside, the Moon has a low topography and thin crust, whereas on the farside, the Moon has a high topography and thick crust. In contrast to the topographic and crustal thickness distribution between the nearside and farside, the Moon has a high concentration of the KREEP material on the nearside, but shows a depletion of the KREEP material on the farside highlands (see Fig. 1). As these asymmetries were likely established soon after the Moon formed, understanding their origin is key for shaping models of the Moon's early evolutionary history (*Shearer et al., 2006*), including the impact cratering process (*Miljkovic et al., 2013; Potter et al., 2015; Zhu et al., 2017*), volcanic activities (*Hess & Parmentier, 2001*), and thermal evolution (*Wieczorek and Phillips, 2000; Zhong et al., 2000; Parmentier et al., 2002; Ghods and Arkani-Hamed, 2007; Zhang et al., 2013; Laneuville et al., 2013; Grimm, 2013*).

Several scenarios have been proposed to explain the possible origins of the nearside-farside asymmetries, including those invoked by external processes such as asymmetric nearside/farside cratering (*Wood, 1973*), ejecta deposition from South Pole-Aitken (SPA) basin (*Zuber et al., 1994*), early inhomogeneous tidal heating (*Garrick-Bethell et al., 2010*), and accretion of a companion moon (*Jutzi and Asphaug, 2011*). Alternatively, thermochemical processes have been suggested as the origin of the nearside-farside asymmetries, such as the magma-ocean convective asymmetries (*Loper and Werner, 2002*), asymmetric mantle contribution (*Wasson and Warren, 1980*), and asymmetric crustal growth (*Ohtake et al., 2012*). In addition, the asymmetric distribution of crustal thickness is coupled with the occurrence of mare basalts over the lunar surface. The majority of mare basalts erupted on the nearside whereas the mare basaltic eruptions on the farside is scarce (*Lucey, et al., 1998*). Two possible scenarios, the degree-1 upwelling (*Zhong et al., 2000*) and downwelling (*Parmentier et al., 2002*) of ilmenite-rich cumulates, have been proposed to explain the nearside-farside asymmetric distribution of mare basalts over the lunar surface. However, as the formation time, bulk composition, crustal structures, and mantle evolution process of the Moon are not completely understood, the origin of the nearside-farside asymmetries remain enigmatic.

Recent observations from the Gravity Recovery and Interior Laboratory (GRAIL) mission suggest a crustal thickness of ∼ 30-40 km on the nearside and ∼ 50-60 km on the farside of the Moon (*Wieczorek et al., 2013*). The analysis of global distribution of major minerals on the lunar surface suggests that the thicker crust on the farside highlands consists of two layers: a primordial anorthositic crust ∼ 30-50 km thick with a more mafic-rich layer ∼ 10 km thick on top (*Yamamoto et al., 2012; Donaldson-Hanna et al., 2014*). Simultaneously, a large area of the low-Ca pyroxene was observed on the nearside by the Kaguya spectral instrument (*Nakamura et al., 2012*). As the low-Ca pyroxene can be formed during an impact by melting a mixture of crust and mantle materials (*Hess, 1994; Warren et al., 1996*), the large area of low-Ca pyroxene on the nearside of the Moon was proposed to have an impact origin (*Nakamura et al., 2012*). According to the most accepted lunar magma ocean (LMO) model (e.g., *Warren, 1985*), the plagioclase feldspar floated to the top of magma ocean that produced the ferroan anorthosite crust and the substantially dense minerals cumulated at the bottom that formed the mafic mantle during the crystallization of magma ocean. The layered crustal structure (e.g., a more mafic-rich layer on top of the primordial anorthositic crust) on the farside (*Yamamoto et al.,* 2012; *Donaldson-Hanna et al.,* 2014), together with the observation of a large area of impact-induced minerals (e.g., low-Ca pyroxene) on the nearside (*Nakamura et al.,* 2012), may indicate that a giant impact on the nearside can explain the observed asymmetries.

A giant impact on the nearside, creating a 3,200-km diameter Procellarum basin (e.g., *Wilhelms and McCauley,* 1971; *Wilhelms,* 1987; *Whitaker,* 1981; see Fig. 1) or a mega basin covering the whole nearside hemisphere (e.g., *Byrne,* 2007), has been proposed more than a decade, but until now this event has not been quantitatively assessed because no obvious structures for this mega basin are observed. The GRAIL data show Bouguer gravity anomalies (*Zuber et al.,* 2013) at the central region of the nearside that have been interpreted as a gigantic tectonomagmatic structure (*Andrews-Hanna et al.,* 2014), rather than a mega basin structure resulting from a giant impact. However, the mega basin structure may not be the same to the surface expression of a typical basin formed later. It is likely that the morphologies of mega basin were heavily modified and possibly even erased due to high internal temperature of the early Moon (*Miljkovic et al.,* 2017), which easily allowed for isostatic relaxation processes (*Freed et al.,* 2014). Therefore, it is highly possible that remnants of a giant impact on the early Moon may be less pronounced and significantly different from the impact basins formed later. To test whether a giant impact on the nearside is a plausible mechanism for the formation of Moon's asymmetries, we performed a systematic numerical modeling study and quantified the outcome of such impact events.

First, we investigate the collision probability of a Ceres-sized impactor (the estimated size range of the impactor in this study) on the early Moon (Section 2). Then we introduce the giant impact simulation (Section 3) and present how we determine the post-impact ejecta and crustal thickness after impact (Section 4). In the results section (Section 5) we present the impact cratering process, the basin size, and the crustal thickness distribution as a result of the giant impact event to see whether such a collision scenario could reproduce the global crustal thickness distribution of the Moon and the layered crustal structures on the farside highlands, as observed; next, we discuss whether such a giant impact event could form the lowlands on the nearside and explain the KREEP distribution on the Moon. Subsequently, we use our giant impact model as an attempt to explain the observed $^{182}$W anomaly in the Moon that is difficult to explain by the Moon-forming impact model (Section 6). Plans for the future studies to further improve the proposed giant impact model for the formation of the nearside-farside asymmetry of the Moon are discussed in Section 7. Finally, we summarize our studies and provide conclusions (Section 8).

## 2. Collision probability of a Ceres-sized impactor on the Moon

The chemical analyses of derivative lunar mantle melts and crustal samples suggest that the Moon is extremely depleted in Highly Siderophile Elements (HSEs) (*Ryder,* 2002; *Walker et al.,* 2004; *Walker,* 2009; *Day et al.,* 2007; *Day et al.,* 2010). The bulk HSE of the Moon, mainly attributed by the post-core-formation late accretion, corresponds to a mass of ∼ 2.1 × $10^{19}$ kg for the chondritic material delivered to the Moon's mantle (*Day et al.,* 2007; *Day et al.,* 2010). For the late accretion, ∼ 60% of the impactor material is thought to be implanted in the Moon and the rest is vaporized and lost into space (*Artemieva and Shuvalov,* 2008). Thus, the usual interpretation for the amount of HSEs is that the Moon accreted ∼ 3.5 × $10^{19}$ kg of chondritic impactor material after its formation (*Morbidelli et al.,* 2012). This amount of material corresponds to ∼ 1.5% of the current mass of the asteroid belt (*Krasinsky et al.,* 2002). Assuming that the size frequency distribution of projectiles that hit the Moon was analog to that of main belt asteroids (*Strom et al.,* 2005), the number of projectiles of any size that hit our satellite is ∼1.5% of the number of corresponding objects existing today in the belt. For a projectile with diameter of ∼ 800-900 km (Ceres-sized impactor), there is only one asteroid in the current main belt asteroid. Therefore, the probability of a Ceres-sized impactor hitting the Moon is ∼ 1.5%. This probability is consistent with that estimated in *Bottke et al.* (2010).

However, the extrapolation backward in time of the decay of lunar bombardment rate over the last 4.1 Gyr (*Neukum and Ivanov,* 1994) suggests that the Moon accreted a mass about 10 to 30 times larger than that inferred from the HSE content (*Morbidelli et al.,* 2012; 2018). These two pieces of information can be reconciled if the lunar HSEs have been sequestered into the lunar core at the end of the crystallization of the LMO and mantle overturn, as proposed in *Morbidelli et al.* (2018) (see also *Rubie et al.,* 2016). In this case, the HSE budget in the Moon would trace only the amount of chondritic material accreted after the LMO crystallization and not the material accreted since the formation of the Moon itself. If the LMO crystallization occurred late, as argued in *Elkins-Tanton et al.* (2011), the difference is relevant. The accretion of 3.5-10 x $10^{20}$ kg of material (i.e., ∼10-30 times larger than usually considered) can be reconciled with the lunar HSE budget if the LMO crystallized about 4.35 Ga ago (*Morbidelli et al.,* 2018). Obviously, if the total mass of projectiles hitting the Moon was ∼ 10-30 times larger than usually considered, the collision probability of a Ceres-sized impactor grows to ∼ 15%-45%. Moreover, it is possible that the size frequency distribution of projectiles was a little different to that of main belt asteroids (*Strom et al.,* 2005), in particular for large projectiles (*Bottke et al.,* 2005; 2010; *Marchi et al.,* 2014; 2018). A slight difference of the size frequency distribution can increase the collision probability of a Ceres-sized impactor to 1 or much bigger.

There is no reason, though, that the size distribution of the projectile that hit the Moon just after its formation was identical to that of main belt asteroids. The lunar projectiles were, in fact, in majority planetesimals left over from the terrestrial planet creation process, namely a distinct reservoir from asteroids, so some differences in size distributions, particular at large sizes, can be expected (*Morbidelli et al.,* 2012; *Morbidelli et al.,* 2018). A different calibration can be obtained by looking at Mars. The formation of the Borealis basin requires a projectile of about 1,500 km in diameter (*Marinova et al.,* 2008). From the leftover of planet accretion, the Moon receives ∼ 1/2 of the impacts of Mars (the ratio of collision probabilities between Mars and the Moon is 2, see *Morbidelli et al.,* 2018). Thus, the impact of an 800-900 km-sized body with the Moon is not unlikely. However, we cannot exclude that the Borealis impact occurred before the Moon-forming event and that no projectiles of comparable sizes were available after the Moon was formed.

## 3. Simulation of giant impact event

We simulate the collision of large bodies with the early Moon using the iSALE shock-physics code (Dellen version, *Collins et al.,* 2004; *Wünnenann et al.,* 2006). We approximate the Moon by a 3,500-km-diameter sphere with a 40-km-thick crust (*Wieczorek et al.,* 2013) on top of a 1,360-km-thick mantle and a 700-km-diameter iron core at the center. We use the Tillotson equation of state for gabbroic anorthosite (*Ahrens and O'Keefe,* 1977), the analytic equation of state for dunite (*Benz et al.,* 1989), and iron (*Thompson and Lauson,* 1972) to calculate the thermodynamic behavior of crustal, mantle, and core material, respectively. The initial densities of these materials are assumed to be 2.9 g cm$^{-3}$ (crust), 3.3 g cm$^{-3}$ (mantle), and 7.8 g cm$^{-3}$ (core). Due to the given size of the impactor with diameter larger than 300 km, we assume that a differentiated structure (*Lee and Halliday,* 1997) has a dunite mantle with density of ∼ 3.0 g cm$^{-3}$ (∼ 70% by mass) and an iron core with density of 7.8 g cm$^{-3}$ (∼ 30% by mass), similar to the average silicate-to-metal ratio for the terrestrial planets. We account for the material strength and dynamic fracturing by a constitutive model according to *Collins et al.,* (2004) and *Ivanov et al.* (2010) in our simulations. We neglect the porosity inside the Moon, first, because that the porosity profiles (*Besserer et al.,* 2014) derived from the current Moon may not represent the characteristics of the early Moon, and second a porous top layer (crust) of the Moon would not affect the basin formation process for a giant impact (*Milbury et al.,* 2015). The material parameters used in the modeling are listed in Table 1.

Since the target's temperature gradient has a significant effect on the formation of large impact basins on the Moon (*Ivanov et al.,* 2010; *Miljkovic et al.,* 2013; *Potter et al.,* 2015), we use two possible thermal gradients for the early Moon in our simulations (see Fig. 2). Both temperature profiles consist of a near surface conductive heat transfer portion (*Potter et al.,* 2012) and a convective portion at depth (*Spohn et al.,* 2001). Thermal profile 1 (TP1) has a crustal gradient of 30 K km$^{-1}$, and the mantle temperature follows an adiabatic gradient (0.5 K/km) at temperatures in excess of 1,300 K (*Freed et al.,* 2014). Thermal profile 2 (TP2) has a crust and upper mantle temperature gradient of 50 K km$^{-1}$; the temperature follows the mantle solidus between a depth of 40-350 km, which would cause partial melting of the upper mantle at this depth range; in the deep mantle (> 350 km depth) the temperature reaches 1,670 K and remains constant (*Potter et al.,* 2012). In contrast with TP1 which represents the thermal gradient of a relatively cold Moon (*Spohn et al.,* 2001), TP2 represents a much warmer thermal profile of the Moon at the time of or before the SPA basin formation (*Potter et al.,* 2012; *Laneuville et al.,* 2013). For the proposed impact scenario, we consider TP2 as the more likely thermal state of the early Moon; however, to have some controls on the effect of the pre-impact temperature on the basin formation we also consider the rather cold TP1 model that may represent the present Moon. As the giant impact was thought to occur prior to the SPA-forming impact, it is possible that the lunar magma ocean had not yet completely crystallized at the time of the giant impact, in which a thin melt-layer might be located between the mantle and the crust. However, as we know very little on the composition and thickness of this melt layer when the giant impact occurred, the temperature variation with depth in this layer is not well constrained. In addition, for the giant impact, the basin forming process and ejecta thickness distribution are not significantly sensitive to the temperature anomaly within a thin layer beneath the lunar crust (*Zhu et al.,* 2017). Therefore, to be conservative, we assume that the temperature of the lower crust and upper mantle, as shown in TP2 profile, follows the solidus to a depth of 350 km. The assumed thermal profiles are comparable to those used in other studies on the formation of lunar basins (e.g., *Potter et al.,* 2013; *Zhu et al.,* 2015), however, other temperature profiles that slightly differ at shallow depth (e.g., *Miljkovic et al.,* 2013) are also possible.

The computational domain covers an area of 5,000 km in the lateral and 9,500 km in vertical direction in all models with a cell size of 10 km in the high-resolution zone (500 x 950 cells). In order to check whether the resolution is sufficient we also carried out some additional runs with a cell size of 5 km in the high-resolution zone. As the results of both resolutions (5-km cell size and 10-km cell size) are very similar on the basin size and ejecta thickness distribution, we use the lower resolution for the majority of models presented here, as the higher resolution models are computationally very expensive.

We set up all the models on a two-dimensional cylindrically symmetric grid, which allows for simulating the vertical impacts only. Two-dimensional models are preferred over three-dimensional models as the latter require far greater computational demands and do not allow for systematic high-resolution parameter studies. In addition, the vertical impacts provide a suitable proxy for impacts over a wide range of impact angles (e.g., *Ivanov et al.,* 2010). As a consequence, the simulation of vertical impact is frequently used for the studies of the large basin formation on the Moon (e.g., *Melosh et al.,* 2013; *Ivanov et al.,* 2010; *Miljkovic et al.,* 2013; *Potter et al.,* 2012; 2013). However, it has been shown that craters tend to become elliptical at large angles with increasing crater efficiency (e.g., *Elbeshausen et al.,* 2013; *Collins et al.,* 2011). For the given size of the impactor, the crater efficiency is relatively low and we cannot rule out that an oblique impact may result in an elliptical crater and an asymmetric crustal thickness. Nevertheless, the 2D vertical impact assumption serves as a first-order estimate to test the feasibility of an impact scenario. After significantly narrowed the range of possible impact parameters, the specific impact scenarios under the consideration of the impact angle will be studied in detail in the future (see the discussions in Session 7 "Further work").

We assume in our models that the gravity at the surface of the sphere representing the Moon corresponds to the average lunar gravitational acceleration of 1.62 m s$^{-2}$. The center of gravity is fixed and located at the center of the lunar sphere. We neglect that this initial gravity field may change slightly due to the redistribution of mass during crater formation. A self-consistent gravity model certainly is more realistic for the modeling of giant impact; however, such simulations are computationally very expensive and do not allow for a systematic parameter study where several hundred models are required. We find that a fixed gravity field cause some unrealistic oscillations of the target around the gravity center, but consider the effect of such oscillations on the basin formation process to be small. Small oscillations in the gravity field do also not change the ejecta trajectories significantly (e.g., *Ivanov et al.,* 2010; *Potter et al.,* 2013). In addition, the similarities on the early stages of basin formation and ejecta emplacement for simulations with self-gravity and central gravity (see Fig. S1 in the supplementary materials) suggest no significant differences on the ejecta trajectories as well. In conclusion we consider the effect of gravity variations as a consequence of crater formation to be small and, for the computational convenience, neglect this effect on the processes we investigate here for the given impact size range. However, with increasing impactor size changes in the gravity field related to the redistribution of mass become gradually more important. Further studies are required to better constrain the size ratio between the impactor and the target where a self-consistent gravity field has to be taken into account (see the discussions in Session 7 "Further work").

For the selected parameters, where the biggest uncertainties exist (e.g., the size and velocity of the impactor), we run models covering the entire plausible range: the impactor diameter is varied between 400 and 1,000 km. The impact velocity is varied between 3 km s$^{-1}$ and 17 km s$^{-1}$ to cover a suitable range of impact velocities for the early Moon (*Marchi et al.,* 2012). In total, we run 240 simulations and each run requires ~ 1 month of computation time. We also run models (120 simulations) with primitive impactors (no iron core), which do not differ too much from models where we include a metal core inside the impactor.

## 4. Ejecta and crustal thickness calculation of giant impact event

As we use the distribution of the ejecta on the surface of the Moon and related changes of the crustal thickness as important constraint of our models in comparison to observations (e.g., crustal thickness model derived from GRAIL gravity data, *Wieczorek et al.,* 2013), we describe here how we derive these parameters, which are not directly provided by our models. We track the ejecta trajectories and their locations using Lagrangian tracers in iSALE. These tracers are initially placed in each computational cell and represent the matter originally in that cell throughout the simulation. When a tracer moves above the pre-impact target surface it is considered as ejecta and its launch time ($t_o$), launch position ($p_o$), and launch velocity ($V$) are recorded (see Fig. 3). We estimate the ejection angle as the angle to the local horizon. For example, for ejecta $a$ with a launch velocity of $V$, we calculate its normal velocity ($V_r$) and tangent velocity ($V_t$), from which the launch angle is estimated to be $\theta$ = atan($V_r/V_t$) (see Fig. 3a). This method has been used previously in several modeling studies on the ejecta distribution (e.g., *Wünnemann et al.,* 2016; *Luther et al.,* 2018). Although the ejecta behave more as a continuous flow with partly fluidized melt, they move along parabolic trajectories and land on the lunar surface with approximately the same velocity as launched. Therefore, it is reasonable to use the hyperparabolic function to estimate their elliptical trajectory for the given ejecta angles and velocities (see Fig. 3b, the elliptical ejecta trajectories for the best-fit model in this work). The flight time ($t_f$), landing position ($p_l$), and landing time ($t_l = t_0 + t_f$) on the lunar surface are also calculated (*Hood and Artemieva,* 2008; *Dobrovolskis,* 1981).

The method to calculate ejecta trajectories separately from the crater formation process by tracers allows to determine the landing site accurately, but not the final deposition. However, it is not trivial to estimate the final location of ejecta because of two reasons. First of all, ejecta do not drop vertically on the ground but strike the surface with the angle and velocity that corresponds to the ejection angle and velocity. Therefore, ejecta may slide along the surface and mix with the local material. The process of entrainment of local material into ejecta blanket upon landing (e.g., *Hörz et al.,* 1983) affects the ejecta stratigraphy significantly and is considered to be important for ejecta deposits several tens to hundred meters thick. For ejecta deposits resulting from large basin forming impacts several kilometers thick we consider this effect to be small and do not take into account here. Second, after the deposition of ejecta, ground motion as a result of the collapse of the transient cavity and subsequent modification process, such as inward slumping of large blocks, takes place and can change the final location of ejecta. The significance of post-emplacement motion depends on the distance to the transient crater rim and is more important close to the rim of the transient crater where modification processes are much more pronounced than at some distance to the crater (*Oberbeck,* 1975). In order to take this effect into account, we firstly determine the flight distance of each tracer along the lunar surface to obtain the ejecta landing site. We then assume that each ejecta moves with the local (target) material around its landing site. For example, for ejecta with a launch position of $\boldsymbol{p_0}$, launch time of $\boldsymbol{t_o}$, flight time of $\boldsymbol{t_f}$, we estimate its landing position of $\boldsymbol{p_l} = \boldsymbol{p_0} + \boldsymbol{p_f}$, here $\boldsymbol{p_f}$ is the flight distance along the lunar surface (see Fig. 3a) (*Dobrovolskis,* 1981). The nearest material on the lunar surface is the closest local material at position $\boldsymbol{p_l}$ at landing time of $\boldsymbol{t_l}$. We track the movement of this local material and consider its final location as the final position of the ejecta hitting on this local material. This method has been used for the study of post-emplacement effect of the ejecta in the modeling of large impact basins on the Moon (*Zhu et al.,* 2015; 2017). The illustration in Fig. 4 shows how we treated the sliding of ejecta once it lands on the Moon.

We calculate the ejecta thickness distribution along the distance from the impact site based on the final position of the ejecta with the effects of post-emplacement motion. We divide the surrounding surface of the basin into discrete rings of 2-arc-degree. The ejecta thickness is calculated from the number of tracers (i.e., mass) located at a given distance from the point of impact up to an arc distance of antipode (180 degrees) assuming the initial density of the material. It is likely that ejecta deposits contain a significant amount of porosity and we may underestimate the ejecta thickness as the assumed density may be too high. However, the given thickness of the ejecta deposits may cause self-compaction, which in turn reduces the porosity. To estimate the porosity of the ejecta deposit is currently not possible and we believe it is the best approach at the current state to assume the largest possible density, which corresponds to zero porosity. The approach to estimate the ejecta thickness as described has been validated against laboratory impact experiments into sand (*Wünnemann et al.,* 2016), where the density problem is not an issue.

Simulations of the ejecta plume (e.g., *Artemieva et al.,* 2013) indicate that the amount of material that is deposited on the ground as a consequence of collapse of the plume and the condensation of vaporized materials produces a very thin layer of dust particle and fine-grained ejecta mixed with the condensation products. We consider this layer to be negligible for the global scale process we are interested in. For this reason and because modeling the fate of the ejecta plume is computationally very expensive we do not account for vaporization of material.

In our models, the post-impact crust is composed of two components: (1) the original pre-impact crustal material and (2) ejecta deposited on top on the original crust. Before impact the original crust of the Moon is assumed to be 40 km thick. As mentioned above, the original crust may be displaced during the basin formation (e.g., crust around the basin rim may slump into the basin interior or flow outwards during the basin-forming process, see section 5.1), we calculate the displaced crustal thickness, similar to the determination of the ejecta thickness, by the number of displaced crustal tracers accumulating into discrete rings of 2-arc-degree width from impact site (0 degrees) up to an arc distance of antipode (180 degrees). As we assume ejecta that lands on the surface to be part of the new crust, the total thickness of the post-impact crust can be calculated by summing up the displaced crustal thickness and the post-sliding ejecta thickness (see Fig. 4). Note, as the ejecta include the material of impactor, and excavated crustal and mantle materials of the target, the ejected mantle materials deposited close to the basin rim that then slumped back into the basin interior are not included in our calculation for the thickness of new crust. It means that the post-impact crust within the basin interior only contains the crustal contribution.

## 5. Results

In total, we carried out 360 simulations of giant impacts to investigate whether such an event could reproduce the crustal thickness asymmetry and two-layer structure on the farside highlands. With the constrains of crustal thickness distribution from the model derived from GRAIL observations, we find that a range of simulations with different combinations of impact velocities and impactor size are in reasonable agreement with the observations. For example, the combinations of impactor diameter ($D_{imp}$) and impact velocity ($v_{imp}$), such as $D_{imp}$ = 500 km with $v_{imp}$ = 15 km s$^{-1}$, $D_{imp}$ = 600 km with $v_{imp}$ =12 km s$^{-1}$, $D_{imp}$ = 720 km with $v_{imp}$ = 6.8 km s$^{-1}$, $D_{imp}$ = 780 km with $v_{imp}$ = 5.2 km s$^{-1}$, $D_{imp}$ = 780 km with $v_{imp}$ = 6.4 km s$^{-1}$, $D_{imp}$ = 780 km with $v_{imp}$ = 7.5 km s$^{-1}$, $D_{imp}$ = 820 km with $v_{imp}$ = 5.2 km s$^{-1}$, and $D_{imp}$ = 820 km with $v_{imp}$ = 5.4 km s$^{-1}$ for a warm Moon (with the temperature profile TP2) reproduce the observed structures. In this section, we report the results of our best-fit model with an impactor with 780 km in diameter (iron core diameter of 200 km) and an impact velocity of 6.4 km s$^{-1}$, which could reproduce not only the observed crustal thickness distribution, but also the two-layer structure on the farside highlands with an ejecta layer ∼ 5-10 km thick on top, as observed. The case of an impactor with 720 km in diameter hitting on a warm Moon with a velocity of 6.8 km s$^{-1}$ also reproduces the observed structures, which is shown in the supplementary materials with other impact combinations (see Fig. S2-S8).

### 5.1 Giant impact cratering process

Figure 5 illustrates the basin-forming process and temperature distribution of the best-fit model. During the basin-forming process, the impactor penetrates into the target, displacing and excavating target material; the floor of the excavated cavity is covered with a thick veneer of impactor and crustal material with a high temperature (Fig. 5b). While the cavity grows, the crater starts to collapse and form a central peak several hundred kilometers high. The outward material flow, formed during the collapse of the central uplift, drags crustal material adjacent to the crater rim into a bulge, giving rise to a thickened crustal region comparable to the farside highlands (Fig. 5c, dashed arrows). Subsequent inward motion of hot material around the basin fills up the excavated cavity, forming a melt pool in the inner basin (~ 2,800 km in radius, see section 5.2). This inward flow transports proximal ejecta back into the basin, erasing typical crater morphological features, such as the elevated rim and thick ejecta blanket adjacent to the rim (Fig. 5d). During the impact process, the impactor's mantle partly mixes with the Moon's crustal material and remains near the surface inside the basin; most of the impactor's iron core (> 85%) mixes with the Moon's mantle beneath the impact site, and only a small fraction merges immediately with the Moon's core (see Fig. 5d).

### 5.2 Basin size for the best-fit model

The giant impact excavates materials down to a depth of ~ 300 km around the impact site (Fig. 6a). The transient crater is defined as the cavity that results from the shock wave-induced excavation flow and marks the end of the so-called excavation stage and the beginning of the modification stage during the impact cratering process (e.g., *Melosh,* 1989). In particular, for basin-sized crater the transient crater does not reflect a certain time in the course of crater formation, but is reached at different place of the transient cavity at different time. For instance, while the gravity-driven collapse occurs at the deepest point of the cavity, the crater continues to grow in diameter. For this reason, it is not straight forward to measure the dimension of the transient crater and different methods have been proposed. *Elbeshausen et al.* (2009) uses the first maximum in crater volume to approximate the transient at a certain point in time. Here we are more interested in the diameter of the transient crater and follow an approach by *Spudis* (1993), where the diameter of the transient crater is given by the radial distance from where material is expelled. We note that estimates of transient crater diameter by both definitions approaches to be equivalent for small craters. However, for the best-fit model of the proposed giant impact, the transient crater radius, defined as the radial distance along the curved surface for the excavation cavity, is ~ 1,800 km (see Fig. 6a), in comparable to ~ 2,000 km estimated by the method of maximum crater cavity (*Elbeshausen et al.,* 2009).

Due to the scale of impact and the thermal conditions considered in this work, the definition of basin rim for the final crater diameter using traditional methods (i.e., a topographic rim) is not trivial. Here, we propose the basin size as the radial distance along the surface from the impact site to a point where more than 90% of the ejecta experience very little (less than a cell size of 10 km) inward sliding after the emplacement within the model time. This definition is plausible to represent the rim of large-scale basins without well-resolved topographic surface expression. The ejecta within the rim slide inward significantly owing to the collapse of the basin walls, whereas most of the ejecta deposited at the rim or further out do not move after landing and thus are not involved in the gravity-driven crater modification process. For the best-fit model, the estimated radius of the final basin is ~ 2,700- 2,800 km. Assuming that the impact site is at 15°N, 23°W, the center of the Procellarum basin (*Wilhelms and McCauley,* 1971; *Wilhelms,* 1987; *Whitaker,* 1981), Fig. 6b shows the boundary of the modeled transient crater (~ 1,800 km) and final crater (~ 2,700 km), from which we find that the boundary of the transient crater is consistent with the extent of the Procellarum KREEP

Terrain (*Jolliff et al.,* 2000), whereas the rim of the final basin (~2,700 km) is located at the boundary between the farside and nearside. The basin size is far greater than the proposed Procellarum basin (~ 3,200 km diameter, *Whitaker,* 1981), but similar to the mega basin proposed on the nearside of the Moon (*Byrne,* 2007).

### 5.3 Post-impact ejecta thickness distribution

The impact velocity of 6.4 km s$^{-1}$ in our best-fit model matches well with the velocities of early impacts on the Moon (*Marchi et al.,* 2012). The low impact velocity results in low ejection velocities, which leads to a relatively small amount of ejecta deposited on the farside highlands. Fig. 7 illustrates the ejecta thickness variation along the radial distance from the impact site before (yellow points) and after (green points) the consideration of ejecta sliding for the best-fit model. The ejecta thickness distribution is determined from the impact site (0 degree) up to an arc distance of the antipode (180 degrees) with each interval of 2-arc-degree width. For the ejecta without considering sliding, the impact produces an ejecta layer more than 110 km at the transient crater rim (yellow points), which decreases to a thickness of ~11 km at an arc distance of 100 degrees from the basin center. The ejecta thickness is 5-10 km at an arc distance of 170 degrees, but increases rapidly to ~ 90 km at the antipode of the impact site, at which the ejecta significantly piles up due to the small area for the 2-arc-degree width.

Assuming that the ejecta, once deposited on the surface, is further transported by material flow, we track this post-excavation movement to obtain the final location of excavated material. The sliding of ejecta changes the material around the rim of the basin (see green points). This is because ejecta near the rim of the basin are carried back into the basin cavity during the modification stage of basin formation (see Fig. 7). When the ejected mantle material that flows back into the basin cavity, it was finally mixed with the local mantle material. The ejected crustal material also flows back into the basin cavity and mixes with local mantle material, however, it will subsequently float to the surface because of its lower density. Note, for the post-sliding ejecta thickness calculation, we do not include the mantle ejecta that flowed back into the cavity. For example, the sharp decrease of the post-sliding ejecta thickness (green points) within 40-90 degrees (the boundary of final crater) in Fig. 7 is because we excluded the ejected mantle material that flowed back into the cavity. However, the decrease of ejecta around the antipode is due to the movement of ejecta in the direction of the impact site where the area of discrete rings is large, resulting in a decrease of the ejecta thickness. The post-sliding ejecta layer is ~ 30 km thick at the transient crater rim, which decreases to a thickness of 10 km at an arc distance of 100 degrees, but retains 5-10 km at an arc distance to 170 degrees from the basin center, corresponding to the farside of the Moon. These ejecta, even partially mixed with the original crust once deposited on the lunar surface, still overlay the original crust, forming the layered crustal structure on the farside highlands with a thickness of ~ 5 – 10 km for the top layer as proposed from the remote sensing observations (*Yamamoto et al.,* 2012; *Donaldson-Hanna et al.,* 2014).

### 5.4 Giant impact reproduces the observed crustal thickness asymmetry

### 5.4.1 The modeled crustal thickness

The modeled crustal thickness (see Fig. 8, orange line), composed of the displaced crust (blue dashed line) and ejecta (green points), varies as a consequence of the outward motion of crust during the collapse of the central uplift (Fig. 5c). It increases from ∼ 30 km at the transient crater boundary to ∼ 50 km at an arc distance of 90 degrees, and reaches a maximum of ∼ 60 km at an arc distance of 140 degrees, then decreases to ∼ 57 km at an arc distance of 180 degrees. Assuming the impact site is at 15°N, 23°W, the center of the proposed mega basin on the nearside of the Moon (*Wilhelms and McCauley,* 1971; *Wilhelms,* 1987; *Whitaker,* 1981; *Byrne,* 2007), Fig. 9a shows the global distribution of the modeled crustal thickness.

### 5.4.2 Comparison with the crustal thickness distribution derived from GRAIL observation

The Moon experienced a long-term impact and volcanism history after the post-giant impact and these activities were thought to have changed the localized crustal thickness distribution, however, the average profile of the observed crustal thickness (Fig. 9b, *Wieczorek et al.,* 2013) will not be significantly changed, therefore, can be directly used to compare with the modeled result. Fig. 9c shows the comparison of the average profiles derived from the modeled crustal thickness (orange, from Fig. 9a) and GRAIL gravity data (black, from Fig. 9b), varied with the radial distance from the impact center. From Fig. 9c, we can find that the giant impact can reproduce the average profile well from the arc distance of 60° (the boundary of transient crater) to 150° (the outer boundary of the SPA basin) within the standard errors. However, the giant impact produces the crustal thickness that reaches a high value in a ring between 120° and 180° from the impact center, whereas the observations show a pronounced crustal thickness between 120° to 150°. We attribute this deviation to the SPA basin with a diameter of ∼ 2,500 km on the farside (see Fig. 9a), which is not considered in our simulation. The SPA basin was thought to have excavated a large amount of the originally thick crust beyond 150° and formed a relatively thin crust (*Potter et al.,* 2012), as observed.

We also use the swathed profiles of the crustal thickness to compare the variation of the localized crustal thickening centered on the farside highlands between the modeled crustal thickness and the crustal thickness derived from the GRAIL gravity data. Swathed profiles, the average data profiles from swaths centered at the farside highlands in the north, northeast, and east directions, can be described by the degree-two harmonics, which was thought to best represent the mean crustal profiles of the Moon and avoid the influences from the other geologic activities (e.g., subsequent impact events and volcanism) (see the details in *Garrick-Bethell et al.* 2010). For each swath, the profile is calculated from transects obtained in two opposite (fit and prediction) directions. The profiles in fit directions were used to check the agreement of the modeled crustal thickness and observed crustal thickness. The profiles in prediction directions are used to test the subsequent effects on the variations of localized crustal thickness on the farside highlands. Table 1 lists the parameters of four swaths used in this work, in which **φ** represents the azimuth of the great circle; **θ** and **ψ** represents the swath width from the original points (longitude and latitude). Each swathed profile is calculated by averaging data profile between two great circle transects emanating from the latitude and longitude. For example, for the fit direction, the azimuths of the first and second great circle transect are defined by the angles of **φ** and **φ+ θ**, separately. The mean profile of the crustal thickness in the fit direction is calculated by averaging great circle transects between **φ** and **φ+ θ**, with each transect having an angular separation of 1 degree. Similarly, the prediction profile is calculated by averaging the data profiles between **φ**+ 180 and **φ+ θ**+ 180. Fig. 9b shows these swaths plotted on the global crustal thickness distribution derived from the GRAIL mission (*Wieczorek et al., 2013*).

We plot four swath profiles from the modeled crustal thickness, together with those derived from GRAIL gravity data (*Wieczorek et al., 2013*) (see Fig. 9d). In Fig. 9d, the profiles in each panel correspond to the swathed profiles derived from the modeled crustal thickness (Fig. 9a) and GRAIL gravity data (Fig. 9b), from which we can find that the modeled crustal thickness reproduces the observed profiles at the farside highlands to an arc distance of 90-105 degrees reasonably well (fit region, represented by the solid green line; correlation coefficient $R^2$ > 0.96). Deviations occur at further distances (> 90 degrees) likely because of effects from subsequent large basin-forming impacts (e.g., Imbrium) that were not considered in this work. In the other direction (< 0 degrees), the modeled crustal-thickness profiles reproduce the observed profiles at the farside highlands to an arc distance of -20 degrees well, but deviations occur at arc distance from -20 degree to -50 degree, in which the observed crustal thickness is much lower than the model predicts (the dashed green line). For swath profiles 3 and 4, these deviations can be attributed to the SPA basin on the farside (see Fig. 9a), which should have excavated a large amount of crustal material (*Potter et al., 2012*). However, the deviations in swath profiles 1 and 2, which have less affect by the SPA and other basins, imply a slightly broad region of the thicker crust predicted on the farside highlands by our impact model, which can be reconciled by a different-sized impactor.

In summary, our results broadly agree with the observed profiles, and therefore, can reproduce the variation of mean crustal profiles of the Moon, which suggests that the giant impact is a possible scenario for the formation of farside highlands and asymmetric crustal-thickness distribution of the Moon.

## 6. Discussion
### 6.1 Impact melt forming the lowlands on the nearside?

As a consequence of shock compression and the subsequent unloading, large amounts of crustal material, as well as the projectile and some parts of the mantle, undergo melting during the impact cratering process. To quantify the melt production, we record the peak shock pressures the material experiences during the passage of shock wave using tracer particles. For the crustal material (gabbro), we assume, for simplicity, the critical peak shock pressures of 56 GPa (*Stöffler, 1972*) and 91 GPa for the softening (the onset of melt) and fully melting and neglect their dependences on the initial temperature because the temperature is low within the crust (*Zhu et al., 2017*). However, for the mantle material, we consider the target's temperature, lithostatic pressure, and density to estimate the critical pressure larger than the solidus and liquidus at different depth. We calculate the post-shock temperatures by relaxing the material to the ambient pressure at the final location of each tracer (*Pierazzo et al.,* 1997). Tracers with post-shock temperatures between the solidus and liquidus were considered to be partially molten, with the fraction of melt varying linearly from zero at the solidus to one at the liquidus (*Potter et al.,* 2013; *Zhu et al.,* 2015). For example, the temperatures of 1,373 K and 2,160 K are considered in our calculation for the softening and melting temperature of the dunite (*Pierazzo et al.,* 1997; *Zhu et al.,* 2017).

The giant impact produces partially molten material (melt fraction > 0) with a total volume of ~ 1.6 x $10^{10}$ km$^3$, where ~ 4.5 x $10^9$ km$^3$ are totally molten (melt fraction = 1). This molten material extends almost entirely over the impact hemisphere (see Fig. 10a). Material with a melt fraction > 0.1 reaches down to the core and extends radially to a distance of 1,600 km (point ***b*** in Fig. 10a), corresponding to the size of mare basalt region (*Jolliff et al.,* 2000) and the putative Procellarum basin (*Wilhelms and McCauley,* 1971) on the nearside (see Fig. 1a and Fig. 10b). According to the modeling of the solidification of impact melt pool within the large impact basins, like Orientale and SPA (*Vaughan et al.,* 2013; *Vaughan and Head,* 2014; *Hurwitz and Kring,* 2014), the impact melt pool will eventually cool and contract on a much longer time scale of ~ 1-10 Ma (*Schwinger and Zhu,* 2018), forming a depression with the similar size of the melt pool. Therefore, it is reasonable to predict that the melt pool produced by the giant impact would form a depression of the same size as the lowlands covering the Moon's nearside (see Fig. 10b) when it eventually cools and contracts. Late-forming impact basins (e.g., Imbrium) can also form a melt pool, but cannot form a depression the same size as the lowlands on the nearside.

Within the basin, the impact melt rocks are mainly composed of target crustal (with some mantle) and impactor mantle rocks, to a radial distance of ~ 900 km (point ***a*** in Fig. 10a). Within this molten zone, the mixture of molten crustal material with mantle material supports the occurrence of olivine and low-calcium pyroxene that likely formed by an impact event (*Nakamura et al.,* 2012). Farther out, up to a radial distance of 1,600 km (point ***b*** in Fig. 10a), the crust does not undergo melting (~ 0 melt fraction), but beyond this distance and up to the boundary of the basin (~ 2,800 km from impact site), the crust is melted again (~ 0.1 melt fraction).

Such a giant impact delivered a large amount of energy to the Moon. *Rolf et al.* (2017) studied the thermal and heat flux anomalies of the Moon as enhanced by a similar sized impact and found that the heat flux anomaly induced by the giant impact last only ~ 100 Myr. However, according to their study, the giant impact can extract more basalts from the mantle within the basin on a long timescale (e.g., ~ 1 Ga, see *Rolf et al.,* 2017), which may contribute to the formation of the majority of mare basalts observed on the nearside of the Moon.

### 6.2 Post-impact KREEP material distribution on the lunar surface

According to the magma ocean model (*Warren,* 1985), the KREEP material was thought to be concentrated in the last magmas to crystallize and sandwiched between the crust and the mantle.

From the remote sensing observations, the KREEP is found predominately on the nearside, but rarely on the farside highlands (*Lawrence et al.,* 1998; *Prettymann et al.,* 2006; *Zhu et al.,* 2013). Several scenarios, including the asymmetric distribution of KREEP material (*Hess and Parmentier,* 2001), inhomogeneous differentiation of the magma ocean (*Loper and Werner,* 2002), and opposite impact ejecta of a companion moon (*Jutzi and Asphaug,* 2011) have been proposed to explain the asymmetric distribution of KREEP material on the Moon. However, its formation is still enigmatic.

In our model, the upper 20 km of mantle is assumed to represent a KREEP layer (*Hess and Parmentier,* 1995) and its location was tracked throughout basin formation. During the basin formation, part of KREEP, at the impact site, is ejected along with crust and mantle material. The ejected KREEP that is deposited close to the basin slumps back into the basin, accumulating at the bottom of, or mixed into, the new crust (see Fig. 11). The other part of KREEP at the impact site is compacted and mixed within the mantle, which may result in the subsequent magmatism widespread on the lunar nearside. In addition, the thick crust around the mega-basin produces heavy stress for the underneath KREEP layer, which may result in the lateral transport, to some extent, and accumulate the KREEP material within the basin (*Manga and Arkani-Hamed,* 1991).

Due to the asymmetric crustal thickness after the giant impact, the depth of the KREEP layer varies between the nearside ($\sim$ 25-30 km) and farside ($\sim$45-55 km). Consequently, the relatively shallow KREEP layer at the nearside could have been excavated and transported more easily to the surface by subsequent basin-forming events (e.g., Imbrium), forming the KREEP distribution with high concentrations on the nearside and low concentrations on the farside (*Lawrence et al.,* 1998; *Prettymann et al.,* 2006). On the farside, significant impacts would be required to excavate KREEP or bring it closer to the surface from its 60-65 km depth. For example, the Orientale basin, with a diameter of 960 km (see Fig. 1), excavated material from a depth of 55 km (*Zhu et al., 2015*).

For the ejecta deposited on the farside highlands, $\sim$ 90% originates from the upper mantle, in which $\sim$ 10% represent the KREEP. Taking an $\sim$ 11 ppm thorium concentration in the vicinity of Imbrium basin rim (*Prettymann et al.,* 2006) as a proxy to represent the KREEP abundance underneath the new crust, the ejecta deposited on the farside highlands have $\sim$ 1 ppm of thorium (10% of ejected KREEP). After the giant impact event, subsequent large impacts and impact gardening on the farside highlands (e.g, *Huang et al.,* 2017) may have further diluted the thorium concentration to < 1 ppm, in agreement with the remote sensing observations (*Prettymann et al.,* 2006). For example, SPA, the largest basin on the farside, was thought to be formed after the proposed giant impact. *Melosh et al.* (2017) (as with *Potter et al.,* 2012) argues that the SPA-forming impact may have excavated the Moon up to a depth of $\sim$ 100 km and produced an ejecta layer with a thickness of several hundreds of meters on the farside highlands. The components of the SPA ejecta deposited on the farside highlands include the lunar crust, mantle, and KREEP materials, with a thorium concentration < 1 ppm (estimated from Fig. 2 in *Melosh et al., 2017*). These ejecta should have mixed and diluted the thorium concentration on the farside highlands produced by the giant impact.

However, we have to note that the initial KREEP concentration underneath the crust before the giant impact is unknown. It is possible that the KREEP may not be evenly globally concentrated underneath the crust at the formation of the giant impact (e.g., *Laneuville et al., 2013*). This is based on the remote sensing observations that the thorium concentration in the vicinity of Imbrium basin (> 10 ppm) in the PKT is greater than the value around the Apollo basin (~ 2 ppm) in the SPA (see Fig. 1c, *Prettymann et al., 2006*). The Apollo basin, ~ 500 km in diameter, should have excavated mantle and, thus also KREEP material (*Potter et al., 2018*). The low thorium concentration at Apollo basin can be only explained by the reduced concentration of KREEP beneath the SPA before its formation (e.g., *Laneuvile et al., 2018*).

### 6.3 Post-impact compositions on the farside highlands

Our model of a giant impact suggests that the post-sliding ejecta with a thickness of 5-10 km on the farside may have formed the lunar highlands. According to our simulation, these ejecta are excavated and expelled from a depth down to ~ 300 km around the impact site (Fig. 6), in which the vast majority (~ 90%) of the ejected materials originate from the upper mantle. These ejecta cover the primordial anorthositic crust and produce a much more mafic-rich layer, forming the two-layer structure on the farside highlands, as suggested by the compositional observations (e.g., *Yamamoto et al., 2012; Donaldson-Hanna et al., 2014*). As the ejecta have high temperature (> 1,800 K, see Fig. 5) and were melted to various degrees along the radial distance from the rim of the basin, they may have differentiated to some extent (*Vaughan et al., 2013; Hurwitz and Kring, 2014*).

The composition of the upper mantle of the early Moon at the time of impact is not well known and assumptions are based mostly on theoretical and indirect considerations. It has been proposed that the upper mantle of the early Moon is dominated by orthopyroxene rather than olivine (*Melosh et al., 2014; Melosh et al., 2017; Lucey et al., 2014*). This assumption is based on the relatively small abundance of olivine in the vicinity of large impact basins (e.g., the SPA basin) that undoubtedly must have excavated mantle material to the surface, spectral data, lunar samples, seismic and petrological studies (*Melosh et al., 2014; Melosh et al., 2017; Lucey et al., 2014; Wieczorek et al., 2006*). An orthopyroxene-rich upper mantle is also supported by modeling of the thermal evolution of the lunar magma ocean (*Klan et al., 2006; Kuskov and Kronrod, 1998; Elkins-Tanton and Bercovici, 2014*). Although the olivine has been observed around the Crisium and Moscoviense basins (*Yamamota et al., 2010*), which may have penetrated through the lunar crust (*Wieczorek et al., 2013; Miljkovic et al., 2015*), the exposures of olivine is thought to originate most likely from the lower crust (*Pieters et al, 2011; Head and Wilson, 1992*). As the distribution of olivine can be assumed to be heterogeneous laterally and vertically within the lower crust (*Martinot et al., 2018*), it cannot be the dominated material of upper mantle of the Moon.

Based on the compositional assumptions of the upper mantle, the proposed giant impact suggests that orthopyroxene-rich mantle material was ejected and deposited at the farside highlands, forming an ejecta layer with a thickness of ~ 5-10 km. According to our simulation, ~ 90% of the ejected materials originate from the upper mantle, which is admittedly much higher than the estimated value of ~ 20% from the remote sensing observations (*Lucey et al., 2014; Crites and Lucey, 2015*). In addition, the ejecta would produce a much higher mafic concentration on the farside highlands than the observations indicated (e.g., ~ 4-5 wt.% of FeO, see *Lucey et al., 2014*). However, the predicted high concentrations of pyroxene and mafic material on the farside highlands do not mean the model results are incompatible with the observations. This is because the proposed giant impact event occurred at an early stage of the evolution of the Moon. After this event the Moon experienced a long-term bombardment history. As a consequence, the ejecta from the proposed impact on the farside highlands would have been covered by ejecta from subsequent giant impacts, such as the SPA basin (*Melosh et al., 2017; Hawke et al., 2003; Yamamoto et al., 2015*). Subsequent large-scale impact basins on the farside highlands (e.g., Birkhoff, Coulomb-Sarton, and Lorentz) would have penetrated through the ejecta layers and exposed the primordial crust to the surface, as it is observed (*Donaldson-Hanna et al., 2014; Ohtake et al., 2009*). The long-term impact gardening (i.e., *Huang et al., 2017*) would have mixed the material on the surface (due to the high degree of complexity this subsequent reworking of ejecta strata has not been considered in this work). These activities may eventually dilute the mafic concentration of materials on the surface of farside highlands to the present-day value (*Lucey et al., 2014*) and produce a relatively low concentration of pyroxene as observed (e.g, *Lucey et al., 2014; Crites and Lucey, 2015*).

The upper mantle of the Moon was thought to have a Mg# (Mg/(Mg+Fe) in mole per cent) of ~ 75-86 to a depth of 270 km, based on the modeling of the thermal evolution of the lunar magma ocean (*Wieczorek et al., 2006; Klan et al., 2006; Kuskov and Kronrod, 1998; Elkins-Tanton and Bercovici, 2014*). The post-impact material on the farside highlands should have a similar Mg#, as estimated from the remote sensing observations (e.g., *Ohtake et al., 2012; Crites and Lucey, 2015*).

### 6.4 Origin of the impactor and implications for the $^{182}$W composition of the Moon

An origin of the Moon's nearside-farside asymmetries by a giant impact implies a substantial addition of mass to the Moon subsequent to its formation. This raises the question of whether this additional mass left an imprint on the Moon's isotopic composition. In this section, we will address this question with a particular emphasis on the $^{182}$W composition of lunar samples, which is sensitive to the late addition of material to planetary mantles.

The Earth and Moon exhibit a striking isotopic similarity for several elements (e.g., O, Ti), which otherwise show large variations among solar system bodies (i.e., asteroids and Mars) (*Wiechert et al., 2001; Young et al., 2016; Zhang et al., 2012*). These isotope anomalies are commonly used as genetic tracers, because they reflect a specific mixture of presolar and nebular components. As such, bodies that formed in different regions of the accretion disk are expected to exhibit distinct isotope compositions. It is therefore remarkable that the Earth and Moon exhibit indistinguishable isotope compositions, although most giant impact models predict that the Moon predominantly consists of impactor material, meaning that the isotopic composition of the Moon should reflect that of the impactor and not the Earth. One way to account for the isotopic homogeneity of the Earth and Moon is to assume that the Moon-forming impactor and proto-Earth both have formed from an isotopically homogeneous inner disk reservoir (*Dauphas, 2014; 2017*). It has also been proposed that the Moon predominantly consists of proto-Earth material (*Cuk and Stewart, 2012; Canup, 2012*), or that there has been post-giant impact equilibration between the Earth and Moon (*Pahlevan and Stevenson, 2007; Lock et al., 2018*).

In the two preferred models considered here, the nearside-farside asymmetry formed by impact of a differentiated body with a total mass of between $7.6 \times 10^{20}$ kg and $9.7 \times 10^{20}$ kg (diameter of between 720 and 780 km). Our model shows that this material would be mixed with a total mass of lunar material of $\sim 3 \times 10^{21}$ kg, and thus contributed $\sim$20–25% of mass to the final mixture between impactor and lunar material. As the proposed impact would have affected a large area on the Moon's nearside, including all Apollo landing sites, it is conceivable that all samples collected at these sites have the isotopic composition of the mixture between the original lunar material and material derived from the asymmetry-forming impactor. The isotopic composition of the Apollo samples, therefore, provides information on the origin of this impactor and on the overall effect of the impact on the Moon's isotopic composition.

The indistinguishable isotopic compositions of the Earth and Moon, as derived from the Apollo lunar samples, imply that the asymmetry-forming impactor had a very similar isotopic composition compared to the Earth and Moon; otherwise the impact would have resulted in an isotopic anomaly in the Moon. For instance, ordinary chondrites, which presumably derive from S-type asteroids (one of the most common asteroid types in the inner solar system) are characterized by a $^{50}$Ti anomaly of $\varepsilon^{50}$Ti$\approx$–0.7 ($\varepsilon^{50}$Ti is the parts-per-10,000 deviation of the $^{50}$Ti/$^{47}$Ti ratio relative to the terrestrial standard) (*Trinquier et al., 2009*). Thus, if the asymmetry-forming impactor had an ordinary chondrite-like isotopic composition, then this would have changed the Ti isotopic composition of the Moon (i.e., the region of the Moon affected by the impact) to $\varepsilon^{50}$Ti$\approx$–0.15 (assuming similar Ti concentrations for impactor and Moon). However, all lunar samples analyzed to date have a common $\varepsilon^{50}$Ti of –0.03±0.04, indistinguishable from the composition of the Earth (*Zhang et al., 2012*). Conversely, to not induce a $^{50}$Ti anomaly in the Moon, the $\varepsilon^{50}$Ti anomaly of the asymmetry-forming impactor must have been within about ±0.25 of that of the Earth. The only known meteoritic material with such small $^{50}$Ti anomalies are enstatite chondrites and aubrites. Although these meteorites do not represent the building material of the Earth and Moon (*Render et al., 2017*), isotopically they are the closest match and, as such, probably formed at a similar heliocentric distance as Earth's building blocks (*Dauphas, 2014; 2017*). The asymmetry-forming impactor, therefore, likely derived from the same region of the accretion disk as Earth's building material, including the Moon-forming impactor.

Whereas the asymmetry-forming impact likely did not induce an isotope anomaly in the Moon for elements like Ti and O, for W isotopes the situation is different. This is because $^{182}$W variations do not result from the heterogeneous distribution of presolar and nebular components, but reflect the chemical fractionation of Hf from W during planetary differentiation and subsequent decay of $^{182}$Hf (half-life = 8.9 Ma) (e.g., *Kleine and Walker, 2017*). Variations in $^{182}$W, therefore, provide a record of the accretion and differentiation history of planetary bodies, and not of their genetic heritage. As the

$^{182}$W composition of the bulk asymmetry-forming impactor likely was chondritic and, hence, different from that of the lunar mantle, this impact likely induced a change in the $^{182}$W composition of the Moon.

We calculated the effect of the impact on the $^{182}$W composition of the lunar mantle, using different $^{182}$W compositions of the impactor's mantle and core. As the impactor composition and differentiation is not *a priori* known, we assumed core formation ages between 1 and 10 Ma after Solar System formation. These timescales are typical for the differentiation of asteroids up to Mars-sized bodies (e.g., *Kleine et al., 2009*) and are, therefore, a reasonable estimate for the differentiation time of the asymmetry-forming impactor. The $^{182}$W compositions of this impactor's mantle and core not only depend on the time of differentiation, but also on the Hf/W ratio of the mantle. This ratio is determined by the metal-silicate partition coefficient *D* for W, which in turn primarily depends on the redox conditions during core formation. We, therefore, calculated the expected $^{182}$W compositions for both a 'reduced' (*D* = 100) and an 'oxidized' impactor (*D* = 5). Fig. 12 shows the effects on the $^{182}$W composition of the lunar mantle (i.e., the area affected by the asymmetry-forming impact) for different impactor compositions and different timescales. The calculations show that this effect strongly depends on the extent to which the impactor core equilibrated with the lunar mantle, and that depending on this effect, the $^{182}$W composition of the lunar mantle can either increase or decrease (here µ$^{182}$W is the part-per-million deviation from the present-day $^{182}$W/$^{184}$W of the bulk silicate Earth). As our simulations suggest that a significant fraction (> 85%) of the impactor core equilibrated with the lunar mantle, the asymmetry-forming impact probably significantly lowered the $^{182}$W composition of that part of the lunar mantle that was affected by this impact. Consequently, the pre-impact $^{182}$W composition of the Moon might have been between ~100 and ~150 µ$^{182}$W (see Fig. 12), implying that prior to the asymmetry-forming impact, the Moon might have had an $^{182}$W anomaly relative to Earth's mantle.

The finding that the Moon originally might have had an $^{182}$W anomaly potentially has important implications for understanding the origin of the Moon. Analyses on lunar samples reveal that the Moon exhibits a ~25 ppm $^{182}$W excess over the present-day bulk silicate Earth (*Kruijer et al., 2015; Kruijer and Kleine, 2017; Touboul et al., 2015*). This $^{182}$W excess is thought to entirely reflect disproportional late accretion to the Earth and Moon. The term 'late accretion' (or late veneer) is commonly used to describe the addition of primitive, broadly chondritic material to the mantles of the Earth and Moon after the Moon-forming impact and the cessation of HSE sequestration into the core (by either metal-silicate differentiation or FeS exsolution during LMO crystallization and mantle overturn). The amount of late-accreted material, as estimated from HSE abundances inferred for each mantle, is much lower for the Moon (~0.02 wt.%) compared to the Earth (0.5–0.8 wt.%) (*Day et al., 2015*). Given this large difference in late-accreted material, mass balance calculations predict that the $^{182}$W composition of the bulk silicate Earth was lowered by ~25 ppm, whereas the lunar $^{182}$W was not changed significantly. As such, the ~25 ppm $^{182}$W difference between the present-day bulk silicate Earth and the Moon can entirely be accounted for by disproportional late accretion (*Kruijer et al., 2015; Kruijer and Kleine, 2017; Touboul et al., 2015*). Consequently, these observations suggest that immediately following the Moon-forming impact, and prior to late accretion, the Earth's mantle and the Moon had indistinguishable $^{182}$W compositions.

However, the $^{182}$W similarity observed between the pre-late veneer Earth's mantle and the Moon is unexpected in the canonical giant-impact model of lunar origin, which predicts a significant $^{182}$W excess in the Moon (*Kruijer and Kleine,* 2017). This conclusion holds regardless of the amount of impactor material in the Moon, meaning that unlike for other isotope systems, the Earth-Moon similarity for $^{182}$W cannot easily be accounted for by making the Moon out of proto-Earth's mantle material. The $^{182}$W data, therefore, suggest that post-giant processes modified the $^{182}$W composition of the Moon, either through Earth-Moon equilibration (*Lock et al.,* 2018) or, as shown in the present study, by the asymmetry-forming impact.

## 7. Further work

It has been argued that most basins on the Moon were formed by the oblique impact. Therefore, a three-dimensional modeling approach using the 3-D iSALE code (*Elbeshausen et al.,* 2009) would be the logical approach for future studies on the formation of the proposed giant impact event and its ejecta distribution. An oblique impact may result in an elliptical crater and an asymmetric crustal thickness in the up-range and the down range direction, depending on the impact angle.

Currently, iSALE 2-D provides greater functionality and can be run at much higher resolution than its 3-D counterpart, thus offering important and detailed insights into basin formation and ejecta distribution. In our simulation, we approximate the gravity field to be static and radially symmetric with the origin at the center of the Moon. Apparently, during the formation of the proposed basin of the given size, the mass distribution temporarily deviates significantly from radial symmetry and therefore the gravity field needs to be adjusted dynamically according to the varying mass distribution. Taking the so-called self-gravity into account would be realistic for the simulation of a giant impact, but is computationally very expensive. The computational demands for systematic study, which requires hundreds of simulations is beyond the computer resources available for this study. However, we consider the proposed impact scenario to be small that self-gravity may be negligible and the gravity field can be considered to be static (central gravity). To test the effect of self-gravity on the cratering process and ejecta emplacement we compare the early stages of basin formation, when mass distribution deviates most from radial symmetry due to the excavation of the transient crater, for a 2D vertical impact model with self-gravity and central gravity (see Fig. S1 in the supplementary materials). The models are almost indistinguishable and, therefore, we assume that the central gravity simplification as used in this study produces the same ejecta layer on the farside highlands as if self-gravity is considered.

The 2D central gravity simulations of vertical impacts in this study are a significant simplification neglecting the effect of obliquity of the impact and the dynamically changing gravity field. However, previous modeling studies (e.g., *Hood and Artemieva,* 2008; *Shuvalov,* 2011) and a simple test comparing crater formation in a dynamically varying gravity field (self-gravity) with a static gravity field (central gravity) show that inside one crater radius, which includes the farside highlands, the ejecta distribution only weakly depends on the impact angle and that self-gravity does not influence at least the early stages of crater formation and ejecta deposition. Although, consider our models as a first-order estimate and late stages processes, such as the gravity-driven crater modification resulting in the final basin morphology, may be more significantly affected by impact angle and self-gravity. However, the given systematic parameters study provides important constraints on the potential formation of the lunar lowlands and highlands as a consequence of a giant impact. More sophisticated 3D self-gravity studies for the proposed impact dimensions are intended as a logical next step, which will be possible upon further code developments of iSALE-3D.

Finally, as the composition of the Moon's upper mantle is still puzzling and little is known for sure about the composition and structure of the lunar farside highlands, further studies are required to investigate the composition of the upper mantle and its relationship to the composition on the farside highlands of the Moon.

## 8. Conclusions

Our results demonstrate, for the first time, that a giant impact on the Moon, like that on Mars (*Marinova et al.,* 2008), could explain the observed asymmetries in crustal thickness, elevation, and surface mineralogy, as the model results match well with observations; the lowlands on the nearside may be a consequence of this giant impact. However, such an impact would produce a basin with a diameter of 5,600 km, covering ∼ 50% of the Moon, far greater than the putative Procellarum basin with a diameter of 3,200 km. Yet, the typical surface expression of this mega basin was erased due to the higher internal temperatures and thus, lower viscosity, of the young planet, resulting in the observed palimpsest-like basins (*Miljkovic et al.,* 2017; *Bottke et al.,* 2015) and a weak basin-like gravity signature (*Miljkovic et al.,* 2017; *Andrews-Hanna et al.,* 2014).

Our model suggests that a significant fraction of the projectile core equilibrated with the mantle. As such, the asymmetry-forming impact may have lowered the µ $^{182}$W value of a large area of the Moon, which would include all of the Apollo landing sites. If more than ∼ 85% of the impactor core equilibrated within the lunar mantle, the pre-impact $^{182}$W composition of the Moon would have been between ∼ 100 and ∼ 150 µ$^{182}$W, implying that prior to the asymmetry-forming impact, the Moon might have had an $^{182}$W anomaly relative to immediate post-giant impact Earth's mantle, as expected in the giant impact model of lunar origin. This model suggests that there might be areas on the lunar farside whose $^{182}$W compositions were less strongly affected. As such, this model might be testable by $^{182}$W measurements on lunar meteorites, some of which probably derive from the Moon's farside.

## References

Ahrens, T. J. & O'Keefe, J. D. (1977). Equations of state and impact induced shock-wave attenuation on the Moon. in *Impact and Explosion Cratering*, edited by D. J. Roddy, R. O. Peppin, and R. B. Merrill. pp. 639–656, Pergamon, New York.

## Acknowledgements

We greatly appreciate reviews from the Editor Steven A. Hauck, Elizabeth Silber, and two anonymous reviewers for insightful reviews that improved the quality of the manuscript. We also thank Jeffery Taylor and anonymous reviewers for valuable comments of the early versions. We thank the developers of iSALE (www.isale-code.de). Tom Davison developed the pySALEPlot tool used in this work. C. Brennecka provided assistance during the paper preparation. The GRAIL crustal thickness data can be accessed online via the GRAIL Crustal Thickness Archive (http://www.ipgp.fr/~wieczor/). The parameters for the iSALE model to reproduce this work are included and illustrated in manuscript. All data to reproduce the figures in this work can be accessed online (https://doi.org/10.5281/zenodo.2644673). This work was funded by the Deutsche Forschungsgemeinschaft (SFB-TRR 170), publication No. 5. M.Z. is partly supported by the Science and Technology Development Fund of Macau (079/2018/A2, 043/2016/A2).

# Figures and Tables

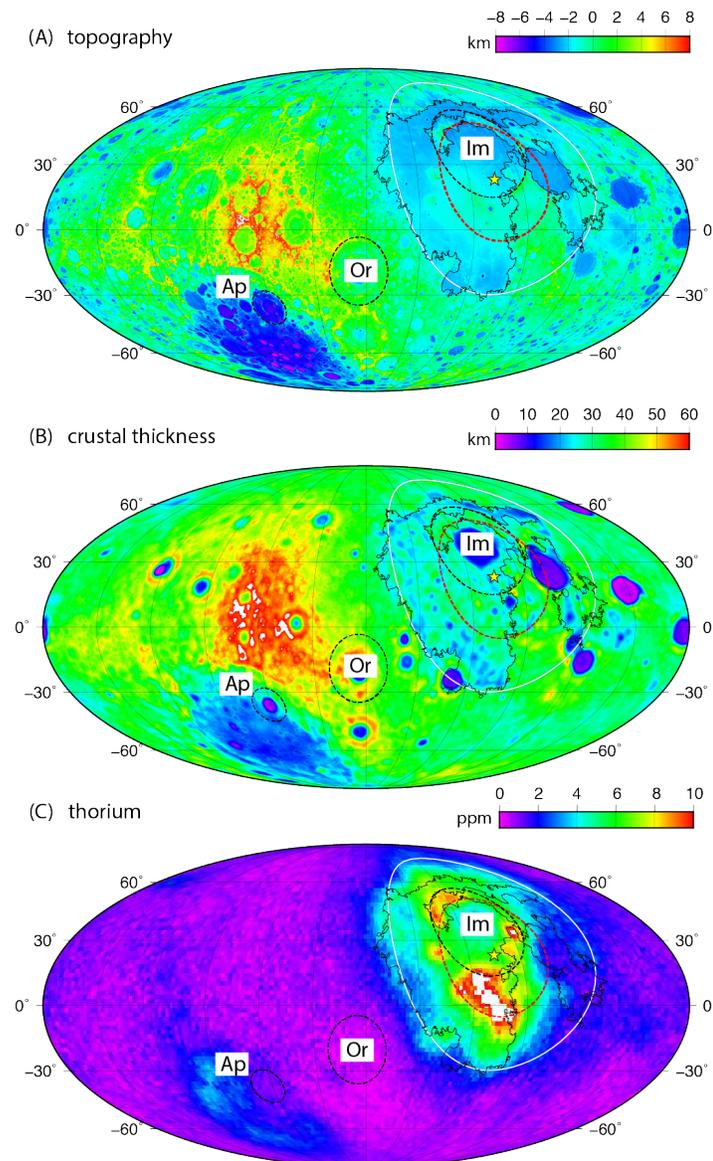

**Figure 1** The topographic (A), crustal thickness (B), and thorium distribution of the Moon show a dichotomy between the nearside and farside. Data are presented in Mollweide equal-area projection centered at 90° W longitude, with nearside on the right of center and farside on the left of center. The star on the nearside represents the center of the proposed Procellarum basin (*Wilhelms,* 1987); the red dashed line and white solid line represent the boundary of transient crater with a radius of 850 km and final basin with a radius of 1,600 km of the proposed Procellarum basin, respectively (*Wilhelms,* 1987). The thin black line on the nearside of the Moon represents the Procellarum KREEP Terrane boundary. The black dashed lines represent the boundary of Imbrium (Im), Orientale (Or), and Apollo (Ap) basin, respectively. The topographic and crustal thickness distribution are derived from the Lunar Orbiter Laser Altimeter (*Smith et al.,* 2010) and GRAIL observations (*Wieczorek et al., 2013*), respectively; the thorium distribution is from the observations of Lunar Prospector gamma ray spectrometer (*Prettyman et al.,* 2006).

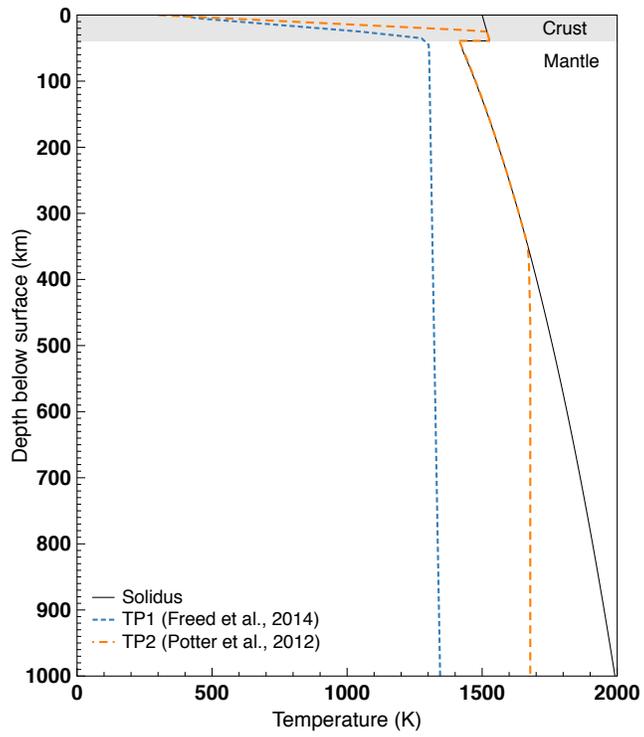

**Figure 2** The thermal profiles of the Moon used in this work. TP1 represents the cold temperature profile of the Moon; TP2 represents the warm temperature profile.

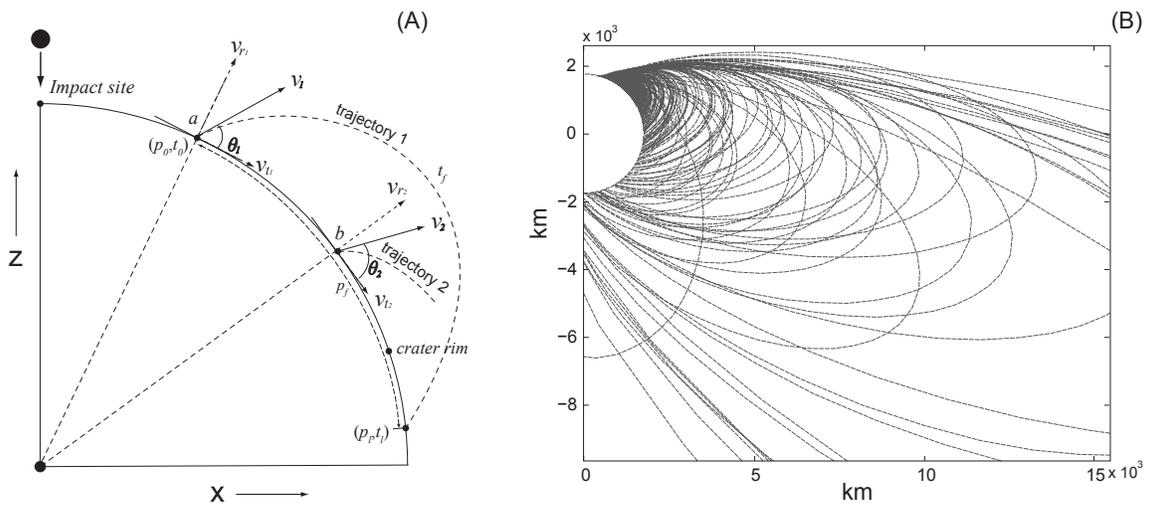

**Figure 3** (A) the launch position($p_0$), velocity($v$), and angle ($\theta$) estimation for the ejecta. The launch time ($t_0$), landing time ($t_l$), flight distance ($t_f$), and landing position of the ejecta are also calculated. Note, $p_f$ defines the flight distance along the lunar surface; (B) the elliptical trajectory for each ejecta for the best-fit model in this work.

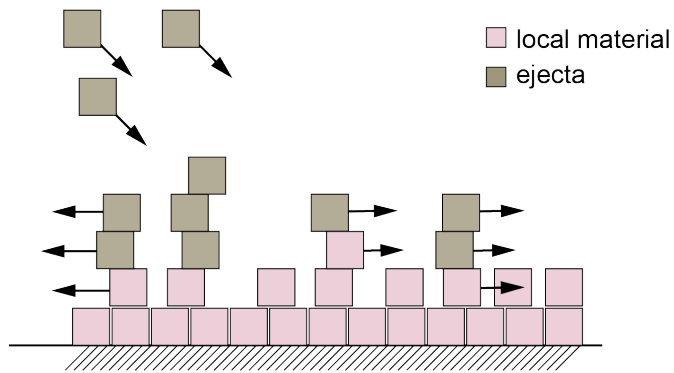

**Figure 4** The carton shows the methods used for the ejecta sliding calculation and the structure of the new crust. The red blocks represent the local crust material over the surface of the Moon. The dark blocks represent the ejecta. The arrows show the directions of movements or the displacements of these blocks. When the ejecta lands on the local material, it follows the movement of the local material. The ejecta, as a new layer covering on the displaced crustal material, and the displaced crust form the post-impact crust.

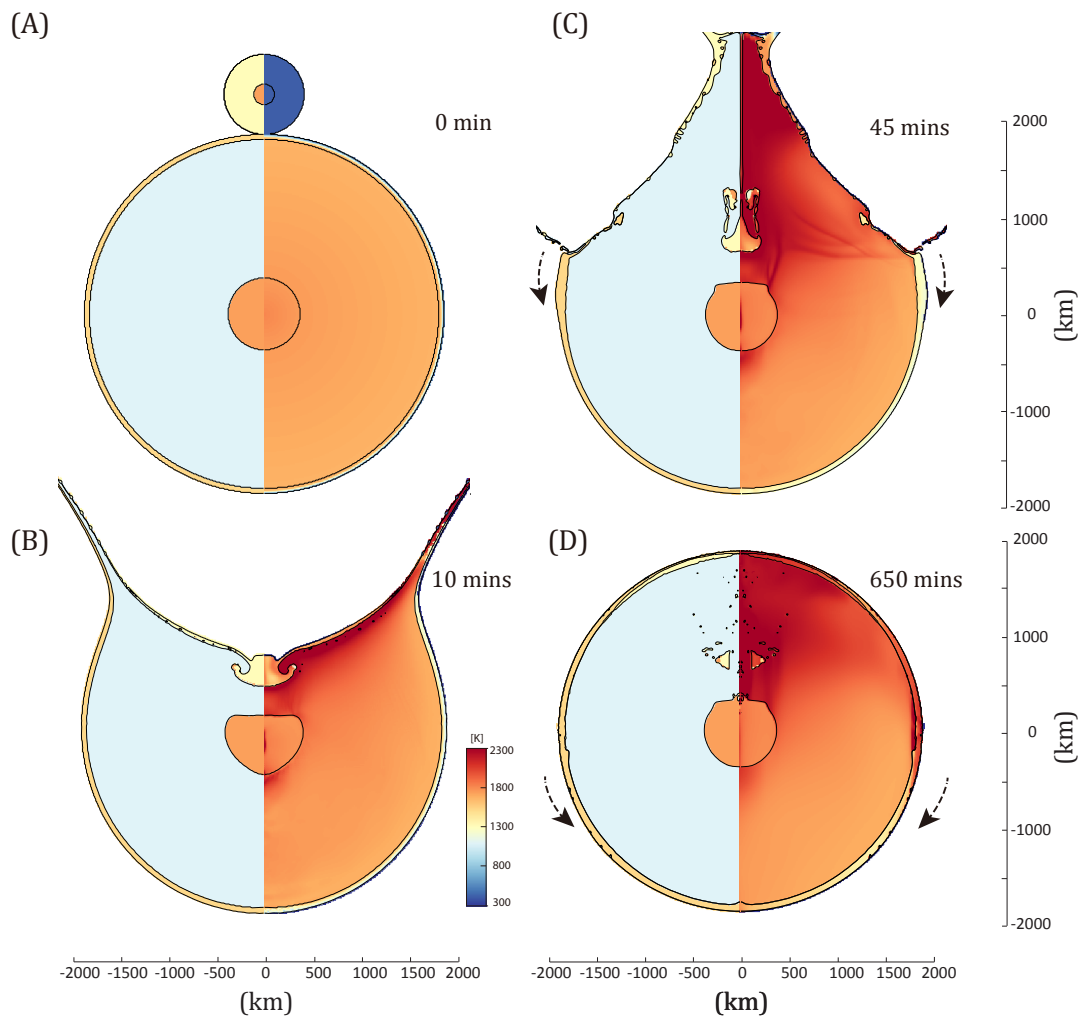

**Figure 5** The basin-forming process for an impactor 780 km in diameter (200 km diameter of iron core) with an impact velocity of 6.4 km s$^{-1}$ using the warmer profile TP2. The Moon was assumed to be a 3,500-km-diameter sphere with 700-km-diameter of iron core. The crustal thickness is assumed to be 40 km. In each panel, the left halves represent the materials used in the model: the gabbroic anorthosite (pale green), dunite (blue), and iron (orange) represent the lunar crust, mantle, and core, respectively. The gabbroic anorthosite (pale yellow) also represents the impactor material. The right halves represent the temperature variation during the impact process. The arrows in (C) and (D) represent the local crustal materials that were displaced and form the new crust together with deposited ejecta.

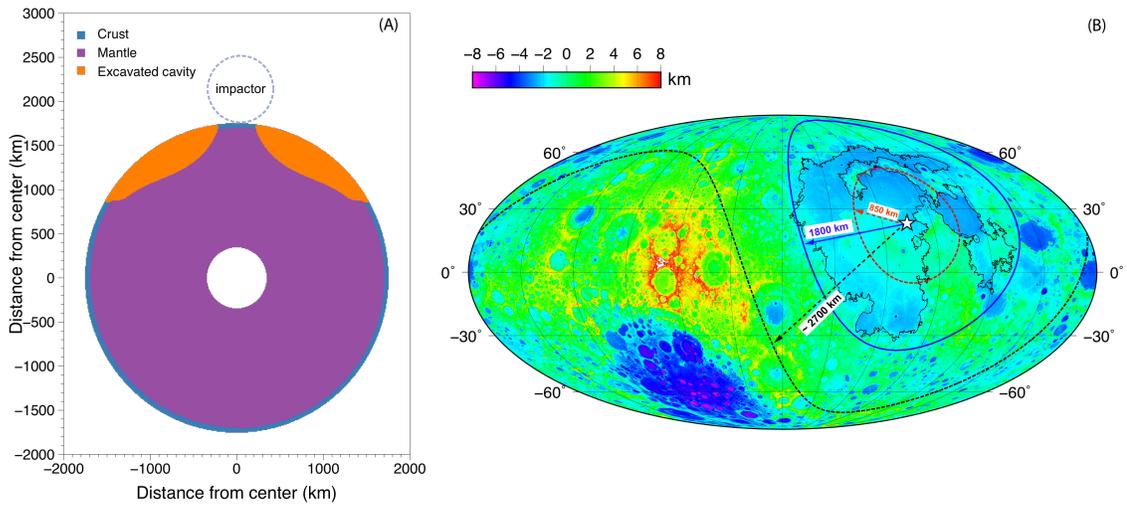

**Figure 6** (A) The ejected material provenance for the giant impact event (780 km diameter impactor with velocity of 6.4 km s$^{-1}$). The crustal thickness was assumed to be 40 km. The excavation cavity is the orange zone. The radius of the excavation cavity is the radial distance along the surface from the impact site to the boundary of excavation cavity. The dashed circle represents the size of impactor. (B) The boundary of transient (solid line) and final crater (black dashed line) for the giant impact event from this work; the transient crater with a radius of 850 km for the proposed Procellarum basin is also plotted as the red dashed circle for the comparison (*Wilhelms,* 1987). Data are presented in Mollweide equal-area projection centered at 90° W longitude, with nearside on the right of center and farside on the left of center. The background is the topographic distribution observed by Lunar Orbiter Laser Altimeter (*Smith et al.,* 2010). The star on the nearside represents the assumed impact site.

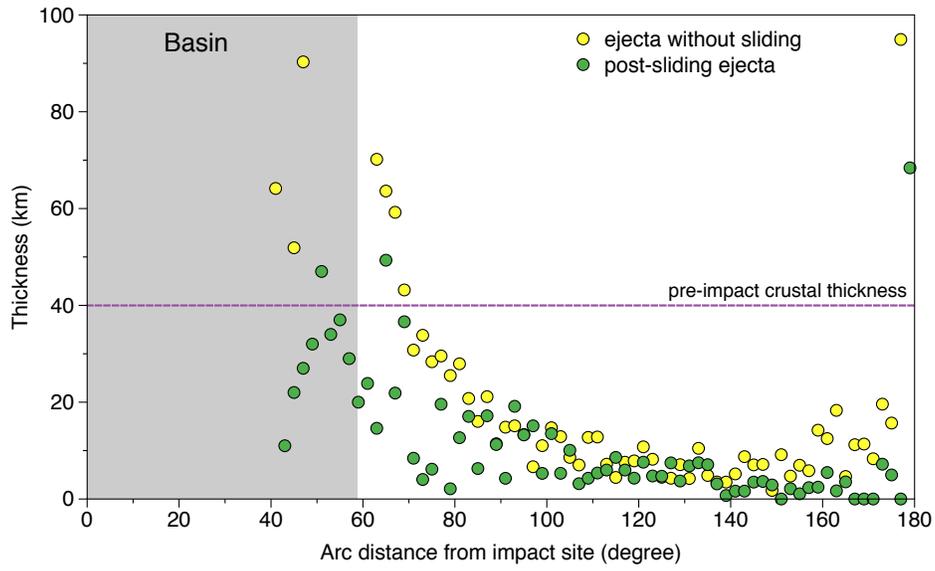

**Figure 7** The ejecta thickness distribution before and after the consideration of ejecta sliding. The gray area represents the extent of the transient cavity. Within the basin (0-90 arc distance), we exclude the mantle material in the post-sliding ejecta.

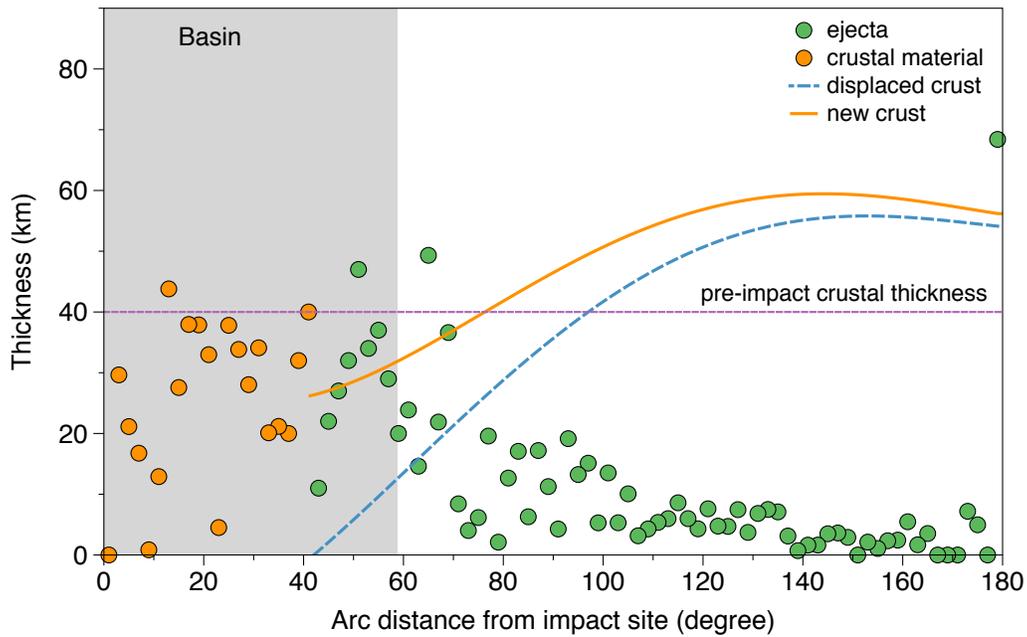

**Figure 8** The ejecta and crustal thickness after post-excavation movement for the impact of a 780 km diameter impactor at 6.4 km s$^{-1}$. The impact site is at 0 degrees. The gray section from 0-60 degrees represents the size of the transient crater (see Fig. 6). Within the transient crater, the ∼ 25-30 km thick crust (average) include the deposited (orange points) and excavated crustal material that underwent post-emplacement transport (green points within 40-60 degrees). The impactor and ejecta thickness distribution are binned into discrete rings of 2-arc degree width from the impact site up to an arc distance of 180 degrees. The orange line represents the impact-derived crustal thickness composed of the displaced crust (blue dashed line) and ejecta (green points), starting from 40 degrees from the impact site. The blue dashed line represents the displaced crust that is originally adjacent to the crater rim and then was dragged by the outward material flow during the impact cratering process.

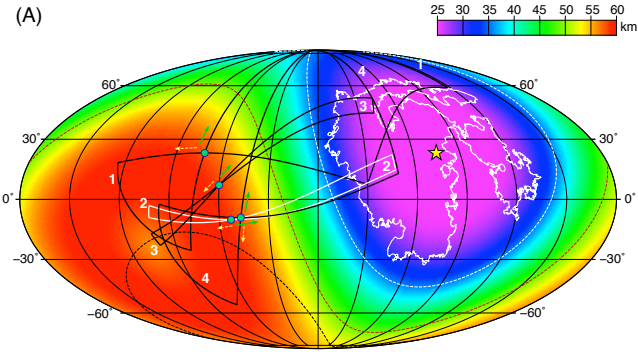

(A)

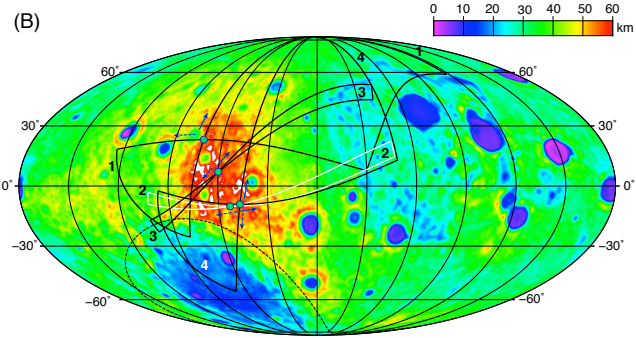

(B)

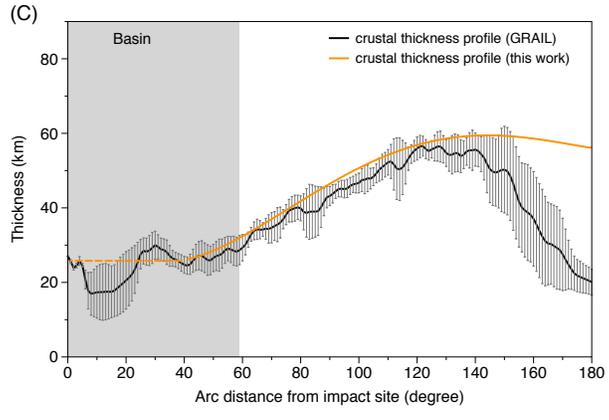

(C)

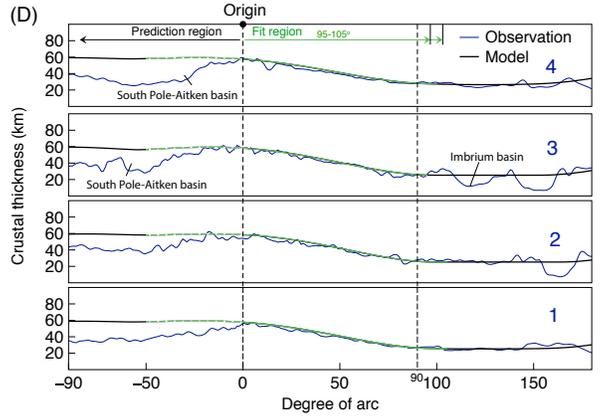

(D)

**Figure 9** (A) map of modeled crustal thickness distribution from the giant impact event; (B) the crustal thickness from GRAIL observations (*Wieczorek et al., 2013*); (C) the comparison of the average crustal-thickness profiles from the modeled crustal thickness (A) and that derived from GRAIL observations (B); (D) the comparison of the average crustal-thickness profiles for swath 1 to 4 derived from both the modeled crustal thickness (A) and the crustal thickness derived from GRAIL observation (B). For (A) and (B), data are presented in Mollweide equal-area projection centered at 90°W longitude, with nearside on the right of center and farside on the left of center. The solid and dashed arrows along each swath represent the fit and prediction direction as shown in Table 2. Points on the swaths represent the centers. In (A), the yellow star represents the impact center and the thin white line represents the Procellarum KREEP Terrane boundary on the nearside; the black dashed line represents the SPA terrain; the white dashed line represents the size of transient crater (r ~ 1,800 km); the red dashed lines represents the final crater (r ~ 2,800 km); the solid and dashed arrow along each swath represent the fit and prediction direction, respectively. In (C), the gray section from 0-60 degrees represents the size of the transient crater; the error bars represent the standard errors for the average crustal-thickens profile. In (D), the thick black and blue lines (from -90° to 180°) represent the mean profiles of swaths derived from the modeled crustal thickness (A) and observed crustal thickness from GRAIL (B), respectively; the green lines (from -50° to 95° or 105°) represent the mean profiles of swaths plotted on (A) and the solid green lines (from 0° to 95° or 105°) represent mean profiles for the fit region.

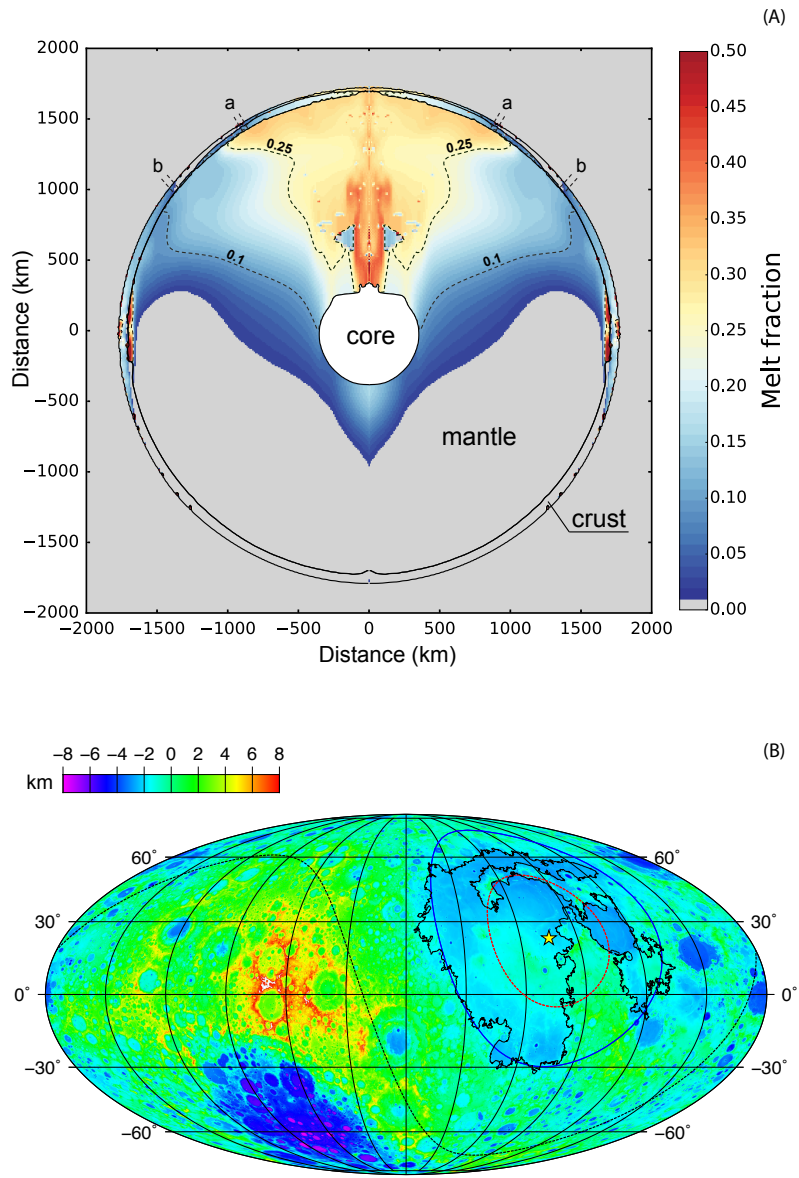

**Figure 10** (A) The melt distribution of the giant impact event. Colored sections represent the molten material; gray sections represent unmolten material (melt fraction of 0). The contours represent the melt fraction with 0.1 and 0.25. Point ***a*** and ***b*** indicates the boundary of the molten and unmolten material of the new crust. The radial distance is ∼ 900 km for point ***a***, and ∼ 1,600 km for point ***b*** from the impact site. The material of the new crust from point ***a*** to ***b*** is not molten. The complete molten material (melt fraction ∼1.0) is located around the boundary of the basin, less than 0.1% of the total of the melt volume. (B) The topographic map of the Moon (left-farside, right-nearside). The black line represents the outline of mare on the nearside. The red dashed and solid blue circle, with a radius of 900 km and 1600 km, represents the boundaries of the molten and unmolten material of the new crust in A (Point ***a*** and ***b***), respectively; the black dashed line represents the boundary of final basin from the modeling. The star on the nearside represents the impact site.

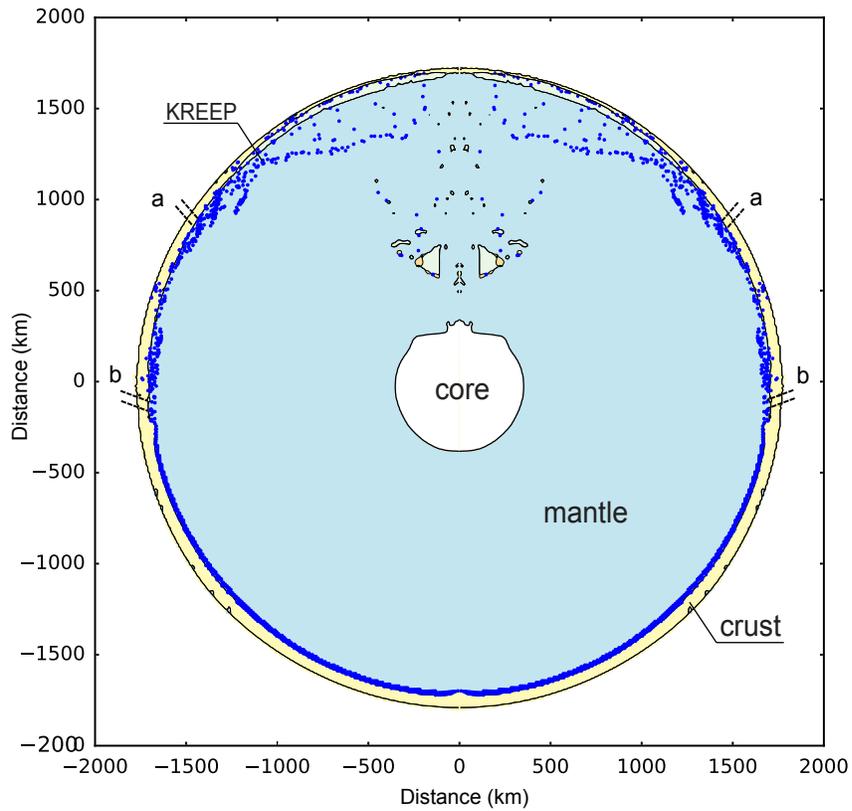

**Figure 11** The post-impact KREEP material distribution. For our model, the global KREEP material was initially resided at the bottom of crust with a thickness of 20 km. The giant impact excavated the crustal, KREEP, and mantle material. The ejected KREEP material, deposited close to the basin rim, slumped back into the basin and was emplaced below crust. Point *a* and *b* represent the boundaries of transient crater (~ 1,800 km) and final crater (~ 2,700 km) from the impact site. See Fig. 6 for the boundaries of transient crater and final crater on the global crustal thickness map.

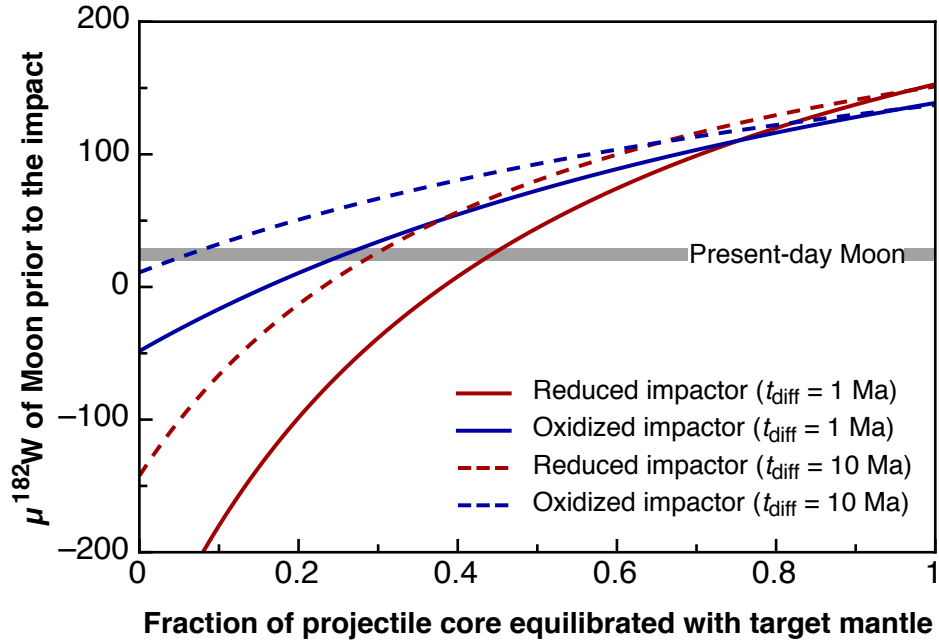

**Figure 12** The predicted µ¹⁸²W of the Moon prior to impact and effects for different impactor compositions and different differentiation timescales after solar system formation. The grey horizontal bar represents the µ¹⁸²W = 24 ± 5 ppm for the Moon (*Kruijer et al., 2015; Touboul et al., 2015*). Calculations assume a mass of the impactor 7.6 × 10²⁰ kg and that the impactor mixed with a mass of 3 × 10²¹ kg of lunar mantle material (within the volume of ∼ 25% melt fraction, see Fig. 10a). Assumed mantle W concentration were 19 ppb (Moon) and 5 (reduced, red lines) and 55 (oxidized, blue lines) ppb in the impactor mantle (see the details in the main text).

**Table 1** Material parameters for iSALE-2D numerical model used in this work

| Description | Crust (target) | Mantle (target/impactor) | Core (target/impactor) |
|---|---|---|---|
| Equation of state | Gabbroic anorthosite Tillotson | Dunite ANEOS | Iron ANEOS |
| Melt temperature at zero pressure (K), $T_{m,0}$ | 1360 | 1436 | 1811 |
| Constant in thermal softening law, $\varepsilon$ | 0.7 | 2.0 | 2.0 |
| Constant in Simon approximation (GPa), $\alpha$ | 4.5 | 1.4 | 107 |
| Exponent in Simon approximation, $c$ | 3.0 | 5.0 | 1.76 |
| Poisson's ratio | 0.25 | 0.25 | 0.30 |
| Cohesion (damaged) (MPa) | 0.01 | 0.01 | |
| Coefficient of internal friction for material (damaged) | 0.6 | 0.6 | |
| Limiting strength at high pressure (damaged) (GPa) | 2.5 | 3.5 | |
| Cohesion (intact) (MPa) | 20 | 50 | 100 |
| Coefficient of internal friction for material (intact) | 1.4 | 1.5 | |
| Limiting strength at high pressure (intact) (GPa) | 2.5 | 3.5 | |
| Minimum failure strain for low pressure states | $10^{-4}$ | $10^{-4}$ | |
| Increase in failure strain with pressure | $10^{-11}$ | $10^{-11}$ | |
| Pressure above which failure is always compressional (MPa) | 0.3 | 0.3 | |
| $\gamma_\eta$ constant | 0.004 | 0.004 | |
| $\gamma_\beta$ constant | 230 | 230 | |
| Peak vibrational velocity as a fraction of the peak particle velocity | 0.1 | 0.1 | |

**Table 2** Swath dimensions from *Gerrick-Bethell* et al (2010)

| Swath | Lat. °N | Lon. °E | °$\varphi$ | °$\theta$ | °$\psi$ |
|---|---|---|---|---|---|
| 1 | 23 | 198 | 0 | 90 | 100 |
| 2 | -10 | 217 | 80 | 10 | 105 |
| 3 | 7 | 210 | 40 | 10 | 95 |
| 4 | -9 | 223 | 0 | 90 | 100 |